\documentclass[aps, pre, twocolumn,superscriptaddress,amsmath,amssymb,floatfix]{revtex4-1} 

\newif\ifpdf\ifx\pdfoutput\undefined\pdffalse\else\pdfoutput=1\pdftrue\fi

\usepackage{graphicx}
\usepackage[usenames]{color} 
\usepackage{comment} 
\usepackage{bm}
\newcounter{abc}

\newcommand{\be}{\begin{equation}} 
\newcommand{\ee}{\end{equation}}
\newcommand{\bea}{\begin{eqnarray}} 
\newcommand{\eea}{\end{eqnarray}}

\begin{document}

\title{Depletion potentials in highly size-asymmetric binary hard-sphere
mixtures: Comparison of accurate simulation results with theory}

\author{Douglas J. Ashton}  
\author{Nigel B. Wilding} 
\affiliation{Department of Physics, University of Bath, Bath BA2 7AY,
United Kingdom} 
\author{Roland Roth}  
\affiliation{Institut f\"{u}r Theoretische Physik, Universit\"{a}t Erlangen-N\"{u}rnberg, Staudtstrasse
7, 91058 Erlangen, Germany} 
\author{Robert Evans}  
\affiliation{H. H. Wills Physics Laboratory, University of Bristol, Royal Fort, Bristol BS8
1TL, United Kingdom}

\begin{abstract}

We report a detailed study, using state-of-the-art simulation and
theoretical methods, of the effective (depletion) potential between a
pair of big hard spheres immersed in a reservoir of much smaller hard
spheres, the size disparity being measured by the ratio of diameters
$q\equiv\sigma_s/\sigma_b$. Small particles are treated grand
canonically, their influence being parameterized in terms of their
packing fraction in the reservoir, $\eta_s^r$. Two specialized Monte
Carlo simulation schemes --the geometrical cluster algorithm, and 
staged particle insertion-- are deployed to obtain accurate depletion
potentials for a number of combinations of $q\leq 0.1$ and $\eta_s^r$.
After applying corrections for simulation finite-size effects, the
depletion potentials are compared with the prediction of new density
functional theory (DFT) calculations based on the insertion trick using
the Rosenfeld functional and several subsequent modifications.  While
agreement between the DFT and simulation is generally good, significant
discrepancies are evident at the largest reservoir packing fraction
accessible to our simulation methods, namely $\eta_s^r=0.35$. These
discrepancies are, however, small compared to those between simulation
and the much poorer predictions of the Derjaguin approximation at this
$\eta_s^r$. The recently proposed morphometric approximation performs
better than Derjaguin but is somewhat poorer than DFT for the size ratios
and small sphere packing fractions that we consider. The effective
potentials from simulation, DFT and the morphometric approximation were
used to compute the second virial coefficient $B_2$ as a function of
$\eta_s^r$. Comparison of the results enables an assessment of the
extent to which DFT can be expected to correctly predict the propensity
towards fluid fluid phase separation in additive binary hard sphere
mixtures with $q\leq 0.1$. In all, the new simulation results provide
the first fully quantitative benchmark for assessing the relative
accuracy of theoretical approaches for calculating depletion potentials
in highly size-asymmetric mixtures.

\end{abstract}

\maketitle

\section{Introduction} 

Much of condensed matter physics and chemistry is concerned with
simplifying the description of a complex many body system by integrating
out certain subsets of the degrees of freedom of the full system. Thus
in treating atomic and molecular solids and liquids one often resorts to
integrating out, usually approximately,  the higher energy quantum
mechanical degrees of freedom of the electrons in order to obtain an
effective interatomic or intermolecular potential energy function that
is then employed in a classical statistical mechanical treatment to
study the properties of condensed phases. Similarly in metallic systems
integrating out the degrees of freedom of the conduction electrons leads
to an effective Hamiltonian for the screened ions or pseudo-atoms. When
one turns to complex, multi-component fluids such as colloidal
suspensions or polymers in solution the basic idea is similar: one
integrates out the degrees of freedom of the small species to obtain an
effective Hamiltonian for the biggest species; for an illuminating
review see \cite{Likos:2001fk}. In this case all species can be treated
classically and the formalism is essentially the famous one of McMillan
and Mayer \cite{McMillan:1945kx} who developed a general theory for the
equilibrium properties of solutions. These authors and many subsequently
recognized that integrating out is best performed when the smaller
species (those constituting the solvent) are treated grand canonically.
Obtaining the full effective Hamiltonian is a tall order. A first step
in any theoretical treatment is to determine the effective potential
between a single pair of the biggest ‘particles’ in a sea of the
smaller species. There is a long history of work in this field.
Different statistical mechanical techniques have been developed to
calculate these potentials for different types of complex fluid.
Well-known examples of effective two-body potentials, forming
cornerstones of colloid science, are the DLVO potential for charged
colloids and the Asakura-Oosawa depletion potential for colloid-polymer mixtures.
Further examples, including polymer systems, are given in
\cite{Likos:2001fk,Belloni2000}.

The problem of determining the effective two-body interaction is particularly
challenging to theory and computer simulation in the situation where
-- and we specialize to a binary fluid -- the bigger particles are much
larger than the smaller ones. Sophisticated theoretical and simulation
techniques, that will provide accurate results for the effective
potential, are only now becoming available. 

Our present focus is on a simple model for a suspension of a binary
mixture of big and small colloidal particles, both species suitably
sterically stabilized, i.e. we consider a highly size asymmetric binary
mixture of hard spheres. This can also serve as a crude model of a
mixture of colloids and non-adsorbing polymer and can  be regarded as a
reference system for a mixture of size asymmetric simple fluids. The
hard sphere mixture is important in that it provides a testing ground
for theories of the effective potential between two big hard spheres
immersed in reservoirs of small ones, with different packing fractions
$\eta_s^r$, and indeed for theories of the full effective Hamiltonian
obtained when the degrees of freedom of the small species are integrated
out fully \cite{Dijkstra1999}. Generally speaking, the more asymmetric
the mixture the more difficult it is to treat both species on equal
footing and the more necessary it is to perform some  integrating out to
obtain an (approximate) effective Hamiltonian. Often this procedure is
only feasible at the level of a Hamiltonian consisting of a sum of
(effective) pair potentials, as obtained by considering a single pair of
the big particles, along  with zero and one body terms \cite{Likos:2001fk,Dijkstra1999}. One of the advantages of the hard
sphere system is that geometrical considerations indicate that three,
four etc body terms become less important as the size ratio
$q=\sigma_s/\sigma_b$ becomes small. $\sigma_s$ and $\sigma_b$ denote
the diameters of the  small and big species, respectively. Thus provided
one can calculate an  accurate effective pair potential, the pair
description alone determines an effective Hamiltonian that should
provide an excellent description of the big-big correlation functions
and the phase behaviour of the binary hard sphere mixture when $q$ is
sufficiently small \cite{Dijkstra1999}. Note that in this paper we consider
additive hard sphere mixtures so that the big-small diameter $\sigma_{bs} =
(\sigma_b+ \sigma_s)/2$. 

Since the studies of the phase behaviour of the hard sphere mixture by
Dijkstra {\em et al} \cite{Dijkstra1999}, whose simulations of an
effective one-component system used a rather crude approximate pair
potential, there have been several new developments in the theory of
effective potentials. Most of these are based on Density Functional
Theory (DFT). It is important to assess whether i) the potentials
derived in recent studies are accurate and ii) use of these might lead
to different predictions for the properties of the mixture.  In order to
make such assessments it is necessary to have accurate simulation
results for the effective potential. Employing state of the art
techniques we provide what we believe are the most accurate results
currently available for $q\leq 0.1$ and packing fractions $\eta_s^r$ up
to $0.35$ and make direct comparisons with the results of theoretical
approaches. The simulation techniques we employ do not allow us to work
at very high values of $\eta_s^r$ but they do allow us to compute
accurate effective potentials, for a range of size ratios, in the regime
of small sphere packing fractions where the putative (metastable)
fluid-fluid phase separation is predicted to occur \cite{Dijkstra1999}.
By calculating the second virial coefficient associated with the
effective potential we make new estimates of the value of $\eta_s^r$
where the onset of this elusive phase transition occurs. Of course  real
colloidal systems may not reach equilibrium on experimental timescales,
particularly if the effective potential exhibits significant repulsive
barriers~\cite{Germain:2009ve}. Nevertheless, knowledge of the underlying phase behaviour is
important for interpreting dynamical observations, such as whether
gelation or glassiness might set in \cite{Poon:1995zr,Lu2008}.

The comparison between DFT results and simulation addresses recent
suggestions \cite{Herring06,Herring07} that no existing theoretical
framework is reliable for calculating effective potentials for small
values of $q$ and physically relevant values of $\eta_s^r$. We examine
and refute these suggestions in the light of our present results.

The effective potential between two big hard spheres takes the form:
\begin{equation}
\phi_{\mathit{eff}} (r_{bb}) = \phi_{bb} (r_{bb}) + W(r_{bb})
\label{eq:effpot}
\end{equation}
where $\phi_{bb}$ is the bare hard sphere potential between two big spheres
and $W$ is the so-called depletion potential. This is attractive for small
separations $r_{bb}$ of the big spheres but decays in an exponentially
damped oscillatory fashion at large separations. The physics of the
attraction is well understood: the exclusion or depletion of the small
spheres as the big ones come close together results in an increase in
free volume available to the small species leading to an increase of
entropy. If this attraction is sufficiently strong it can give rise to
fluid-fluid phase separation. Such a phase transition is driven by
purely entropic effects: recall that all the bare interactions in the
mixture are those of hard spheres. Of course the concept of an attractive depletion potential between
colloids dispersed in a solution of non-absorbing polymer, or other
depletants has a long history. The recent book by Lekkerkerker and
Tuinier \cite{Lekkerkerker:2011} describes this and the general
importance of depletion interactions in colloidal systems. 

We have emphasized that the effective pair potential is a key ingredient
in an effective Hamiltonian description of the mixture. However, this
object is also important in its own right since it can be measured
experimentally for colloidal systems using various techniques. More
specifically, the effective potential between a single colloid, immersed
in a suspension of small colloidal particles or non-adsorbing polymer,
and a flat substrate has been measured; see for example
\cite{Rudhardt1998} and the comparisons  made between DFT results and
experiment \cite{BECHINGER99}. Crocker {\em et al} \cite{Crocker1999}
measured the effective potential between two big PMMA particles immersed
in a sea of small polystyrene spheres and observed damped oscillations
at high small sphere packing fractions. Subsequently comparisons were
made with DFT results \cite{Roth2000a}. Ref.~\cite{Kleshchanok:2008vn}
provides a recent review of direct experimental measurements of
effective interactions in colloid-polymer systems.

\subsection{Previous Simulation Studies}

\label{sec:sims} 

In general, the task of accurately measuring effective potentials in
highly size-asymmetrical fluid mixtures using traditional simulation
methods such as molecular dynamics (MD) or basic Monte Carlo (MC) is an
extremely challenging one. The difficulty stems from the slow relaxation
of the big particles caused by the presence of the small ones.
Specifically, in order for a big particle to relax,  it must move a
distance of order its own diameter $\sigma_b$. However, for small size
ratios $q$, and even at quite low values of $\eta_s^r$, very many small
particles are typically to be found surrounding a big particle and these
hem it in, greatly hindering its movement. In computational terms this mandates a very small
MD timestep in order to control integration errors, while in MC, a very
small trial step-size must be used in order to maintain a reasonable
acceptance rate. Consequently, the computational investment required to
simulate highly size asymmetric mixtures by traditional means is
(generally speaking) prohibitive at all but the lowest packing fractions
of small particles. 

Owing to these difficulties, most previous simulation studies of hard sphere mixtures
\cite{Biben1996,Dickman1997,Goetzelmann1999,Herring06,Herring07}
have adopted an indirect route to measuring effective potentials in the
highly size asymmetrical regime $q\leq 0.1$ based on measurements of
interparticle {\em force}.  The strategy rests on the observation that
the force between two big particles can be expressed in terms of the
contact density of small particles at the surface of the big ones
\cite{Attard1989,Dickman1997}. By measuring this (angularly dependent)
contact density for fixed separation $r_{bb}$ of the big particles and
repeating for separations ranging from the minimum value
$r_{bb}=\sigma_b$ to $r_{bb}=\infty$, one obtains the force profile
$F(r_{bb})$. This can in turn be integrated to yield an estimate of the
depletion potential. Generally speaking, however, the statistical
quality of the data obtained via this route seems typically quite low,
particularly at small $q$ and high $\eta_s^r$. This presumably reflects
the difficulties of measuring contact densities accurately and the
errors inherent in numerical integration.

Only a few studies have tried to measure the depletion potential
directly for $q\le 0.1$ (see ref.~\cite{Malherbe2001} for a hard sphere
study  and refs.~\cite{Liu2005,Barr2006} for more general potentials).
In common with the present work, these studies deployed a cluster
algorithm (to be described in sec.~\ref{sec:gca}) to deal
efficiently with the problem of slow relaxation outlined above. However,
they treated the small particles canonically rather than grand
canonically, which complicates comparison with theoretical predictions
which are typically formulated in terms of an infinite reservoir of
small particles \footnote{Though see \protect\cite{Goetzelmann1999} for
a grand canonical study at $q=0.2$}. Furthermore it seems that no
previous simulation studies have discussed (in any detail) finite-size
effects in measurements of depletion potentials, the role of which we
believe to be particularly significant at large size asymmetries.
Consequently, while previous work has evidenced good qualitative
agreement between simulation and theory, there is to date a lack of data
from which one can make confident comparisons between the various
theoretical approaches. This is remedied in the present work.

\subsection{Previous Theoretical Studies}

There are many of these and they are based on a variety of techniques.
Integral equation treatments abound and these are summarized nicely in
the recent article by Bo\c{t}an {\em et al} \cite{Botan2009}.  A related
rational function approach was used recently by Yuste {\em et al}
\cite{Yuste2008}.  Tackling highly asymmetric mixtures via integral
equation methods, where one treats both species on equal footing, is
notoriously difficult and making systematic assessments of the
reliability of closure approximations is not straightforward and, of
course, requires accurate simulation data. 

Density functional (DFT) treatments are arguably much more powerful.  There
are several different ways of calculating the effective (depletion)
potential between two big hard spheres in a reservoir of small hard
spheres – or more generally of calculating the effective potential
between two big particles in a reservoir of small ones with arbitrary
interactions between {\it bb, bs} and {\it ss}.  The first method is to fix the
centres of the big ($b$) particles a distance $r_{bb}$ apart and then compute
the grand potential of the small ($s$) particles in the external field
of the two fixed $b$ particles as a function of $r_{bb}$ for a given size
ratio $q$ and a reservoir packing fraction of  $\eta_s^r$.  This method
requires only a DFT for a single component fluid, the small particles.
The big particles are fixed so they simply exert an external potential
on the small ones.  DFT for one-component hard spheres is very
well-developed; very accurate functionals exist and these are suitable
for treating the extreme inhomogeneities that arise for small size
ratios $q$. It follows that this ‘brute force’ method should be rather
accurate. Its drawback is that the density profile of the small
particles has cylindrical symmetry requiring numerically accurate
minimization of the free energy functional on a two-dimensional grid. 

Goulding \cite{Goulding:2001fk,Goulding2000} performed pioneering brute
force calculations using the Rosenfeld fundamental measure theory (FMT)
\cite{RosenfeldPRL1989} for a system with $q=0.2$ and packing fractions
$\eta_s^r$ up to $0.314$. This method has been refined recently by Bo\c{t}an
{\em et al} \cite{Botan2009} who employed various hard-sphere functionals for
more asymmetric systems and higher values of $\eta_s^r$. These authors,
see also Oettel {\em et al} \cite{Oettel2009}, also used DFT to calculate the
depletion force directly using the formula due to Attard
\cite{Dickman1997,Attard1989} that relates the force to the density
profile of small spheres in contact with a big sphere. Once again the
density profile has cylindrical symmetry and careful numerical methods
are required. 

A popular DFT method for hard-sphere systems is based on what has
become known as the insertion trick or insertion method
\cite{Goetzelmann1999,Roth2000a}.  This is a general procedure, see
Sec.~\ref{sec:DFT}, for calculating the depletion potential between a
big particle and a fixed object, e.g. a wall or another big particle,
immersed in a sea of small particles.  The advantage of the method is
that one requires only the equilibrium density profile of the small
species in the external field of the isolated fixed object and this
profile clearly has the symmetry of the single fixed object.  For the
case of two big spheres the profile $\rho_s(r)$ has spherical symmetry.
The disadvantage is that the theory requires a DFT for an asymmetric
mixture, albeit in the limit where the density $\rho_b$ of the big
particles approaches zero:  $\rho_b\to 0$.  For hard-sphere mixtures the
insertion method is straightforward to implement 
and \cite{Roth2000a} provides a series of comparisons, using the Rosenfeld FMT
\cite{RosenfeldPRL1989}, with the simulation data that existed in 2000.

Further comparisons between results of the DFT insertion method and
simulation studies were made in Refs.~\cite{Herring06,Herring07,
Botan2009,Oettel2009}. In the present paper we seek to make more
quantitative comparisons, taking into consideration the improved
accuracy of our new simulation results and the availability of improved
DFTs for hard sphere fluids. 

There is a further theoretical approach to calculating depletion
potentials developed very recently in Ref.~\cite{Oettel2009} and
employed subsequently in Ref.~\cite{Botan2009}. This approach is based on
morphological (or morphometric) thermodynamics \cite{Konig:2004uq}. The
basic idea is that the depletion potential is (essentially) the
solvation free energy of the dumbbell formed by the two big spheres and
that this quantity can be separated into geometrical measures, namely
the volume, surface area and integrated mean and Gaussian curvatures.
The coefficients of these measures are geometry independent
thermodynamic quantities, i.e. the pressure, the planar surface tension
and two bending rigidities all of which can be obtained from simulations
or from DFT calculations of the single component fluid performed for a
simple geometry. 

The paper is arranged as follows:  Section~\ref{sec:simmeth} describes
the grand canonical simulation methods that we have employed for
determining the depletion potential between two big spheres for size
ratios $q$ from $0.1$ to $0.01$. In Section~\ref{sec:theory} we
summarize briefly the DFT insertion method and the three hard sphere
functionals that we employ in our present calculations. We also discuss
some of the limitations of the parameterization of the depletion
potential introduced in \cite{Roth2000a}. Some details of the
morphometric approaches are also given here. Results are presented in
Section~\ref{sec:results}. As a test of our simulation method we
determine the depletion potential for two big hard spheres in a solvent
of non-interacting point particles that have a hard interaction with the
big spheres. For this case the depletion potential is known
analytically--it is the venerable Asakura-Oosawa potential
\cite{Asakura1954,Asakura:1958uq} of colloid science. For the additive
binary hard sphere case we make comparisons between results of
simulation, DFT insertion method, the morphometric approach and the
Derjaguin approximation \cite{Goetzelmann1998} for the depletion
potential. We also compare simulation, DFT and morphometric results for the associated
second virial coefficient $B_2(\eta_s^r)$. The latter provides a
valuable indicator of the propensity of the bulk binary mixture to phase
separate into two fluid phases \cite{Roth2001,Largo2006}. We conclude in
Section~\ref{sec:discuss} with a discussion.

\section{Simulation Methods}
\label{sec:simmeth}
\subsection{Geometrical cluster algorithm}
\label{sec:gca}

An efficient cluster algorithm capable of dealing with hard spheres mixtures was
introduced by Dress and Krauth in 1995 \cite{Dress1995}.  It was
subsequently generalized to arbitrary interaction potentials by Liu and
Luijten \cite{Liu2004,Liu2005} who dubbed their method the Geometrical
Cluster Algorithm (GCA). Here we describe the application of the GCA to
a size asymmetrical binary mixture of hard spheres.

The particles comprising the system are assumed to be contained in a
periodically replicated cubic simulation box of volume~$V$.  The
configuration space of these particles is explored via cluster updates, in
which a subset of the particles known as the ``cluster'' is displaced
via a point reflection operation in a randomly chosen pivot point. The
cluster generally comprises both big and small particles and by virtue
of the symmetry of the point reflection, members of the cluster retain
their relative positions under the cluster move. Importantly, cluster
moves are rejection-free even for arbitrary interparticle interactions
\cite{Liu2005}. This is because the manner in which a cluster is built
ensures that the new configuration is automatically Boltzmann
distributed.
 
For hard spheres, the cluster is constructed as follows: one of the
particles is chosen at random to be the seed particle of the cluster. 
This particle is point-reflected with respect to the pivot from its
original position to a new position.  However, in its new position, the
seed particle may overlap with other particles. The identities of all
such overlapping particles are recorded in a list or ``stack''. One then
takes the top-most particle off the stack, and reflects its position
with respect to the pivot. Any particles which overlap with this
particle at its destination site are then added to the bottom of the
stack. This process is repeated iteratively until the stack is empty and
there are no more overlaps. 

In this work we shall be concerned with measurements of the radial     
distribution function $g_{bb}(r)$ for a system containing a pair of big
particles among many small ones. To effect this measurement we modify
the GCA slightly as follows: we choose one big particle to be the seed
particle, which we place randomly within a shell $\sigma_{bb} < r <
L/2$, centred on the second big particle, with $L$ the linear box
dimension. The location of the pivot is then inferred from the old and
new positions of the seed particle. Thereafter clusters are built in the
standard way. This strategy ensures that we efficiently sample separations of
the big particles that lie in the range for which $g_{bb}(r)$ can
sensibly be defined for hard spheres in a cubic box.

Small particles are treated grand canonically in our simulations. In
practical terms this means that in parallel with the cluster moves, we
implement insertions and deletions of small particles, subject to a
Metropolis acceptance criterion governed by an imposed chemical
potential. The choice of chemical potential controls the reservoir
packing fraction of small particles.

For the systems of interest in this work, we find that the GCA is
efficient for reservoir packing fractions $\eta_s^r\lesssim 0.2$. Above
this value one finds that practically all the particles join the
cluster, which merely results in a trivial point reflection of the
entire system. For single component fluids this problem can be
ameliorated by biasing the choice of pivot position  to be close to
the position of the seed particle \cite{Liu2005}. Doing so has been
reported to extend the operating limit to $\eta_s^r\simeq 0.34$.
However for the case of highly asymmetrical mixtures we find that this
strategy does not significantly decrease the number of particles in the
cluster because as soon as a second big particle joins the cluster and is
point reflected it causes many overlaps with small particles.

\subsection{Staged insertion algorithm}
\label{sec:staged}

Our second MC approach for obtaining effective potentials for size
asymmetrical mixtures is based on the staged insertion of a big
particle \cite{Nezbeda1991,Attard1993,Wilding1994a,Ashton:2011fk}. The
method involves first fixing one big particle at the origin and then
sampling the free energy change associated with inserting a second big
particle at a prescribed distance $r_{bb}$ from the origin. Essentially
this amounts to an estimation of the chemical potential of the second
particle $\mu_{ex}(r_{bb})$. As such our approach is close in spirit to
one proposed very recently by Mladek and Frenkel \cite{Mladek2011},
although their implementation did not employ staged insertion and was
therefore restricted to low density systems or those interacting via
very soft potentials.

The effective potential for big particle separation $r_{bb}$ is simply

\be
W(r_{bb})=\mu_{ex}(r_{bb}) + C
\label{eq:eff_mu}
\ee
where the additive constant 
\be
C=-\lim_{r_{bb}\to\infty}\mu_{ex}(r_{bb})
\ee
can be determined as the excess chemical potential of a single big
particle in the reservoir of small particles. To estimate
$\mu_{ex}(r_{bb})$ we follow the strategy described in
Ref.~\cite{Wilding1994a}. In outline, one imagines that the second big
particle can exist in one of $M$ possible `ghost' states in which  it
interacts with a small hard particle (a distance $r_{bs}$ away) via the potential

\[
\beta\phi_g^{(m)}(r_{bs})=-[1-\Theta(2r_{bs}-\sigma_b)]\ln\lambda^{(m)}\:.
\]
Here $m=1\ldots M$ (an integer) indexes the ghost states, while the associated
coupling parameter $0\le \lambda^{(m)}\le 1$ controls the strength of
the repulsion between the big particle and the small one. Owing to the step function
$\Theta$, the potential is uniformly repulsive over the volume of the
big particle, and zero elsewhere. Moreover, for $\lambda^{(m)}>0$ the
repulsion is {\em finite} so that overlaps between small particles and the big
one can occur. If we denote by $N_o$ the number of such overlaps at any given
time, then the configurational energy associated with the ghost big particle
is

\be
\beta\Phi_g^{(m)}=-N_o\ln\lambda^{(m)}\:.
\label{eq:energy}
\ee

Clearly for $\lambda^{(m)}=0$, the big particle acts like a normal
hard sphere, while for $\lambda^{(m)}=1$ there is no interaction and 
the big particle is effectively absent from the system.  To span this
range we set the extremal states $\lambda^{(1)}=0$ and
$\lambda^{(M)}=1$, and define some number of intermediate states that
facilitate efficient MC sampling over the range $m=1\ldots M$, i.e. that
permits the ghost particle to fluctuate between being a real hard sphere
and being absent. The measured value of the relative probability of
finding the system in these extremal states yields the excess chemical
potential:
 
\be
\mu_{ex}(r_{bb})=\ln\left[\frac{p(\lambda^{(M)})}{p(\lambda^{(1)})}\right]\:.
\label{eq:probs}
\ee

Now since $W(r_{bb})$ is spherically symmetric, it can be estimated from
Eqs.~\ref{eq:probs} and \ref{eq:eff_mu}, by measuring $\mu_{ex}(r_{bb})$ for
values of $r_{bb}\ge\sigma_b$ along a 1D grid. Moreover since each such
measurement is independent of the others, the approach is trivially
parallel and thus can be effectively farmed out on multi-core
processors.

Details of a suitable Metropolis scheme for sampling the full range of
$m=1\ldots M$ have been described in detail previously
~\cite{Wilding1994a,Ashton:2011fk}. The basic idea is to perform grand
canonical simulation of the small particles, supplemented by MC updates
that allow transitions $m\to m\pm 1$ for the ghost big particle. These
transitions are accepted or rejected on the basis of the change in the
configurational energy Eq.~\ref{eq:energy}. However, for this strategy
to realize the aim of sampling the relative probability of the extremal
states, it is necessary to bias the transitions such as to ensure
approximately uniform sampling of the $M$ ghost states. This is achieved
by determining a suitable set of weights which appear in the MC
acceptance probability \cite{Lyubartsev1992}. 

Additionally it is important to choose sufficient intermediate states
and to place them at appropriate values of $\lambda$ such that
transitions $m\to m\pm 1$ are approximately equally likely in both
directions and have a reasonably high rate of acceptance. To achieve
this we perform a preliminary run in which we consider a single big
ghost particle in the reservoir of small ones. We first define $M=1000$
values of $\lambda$ in the range $(0,1)$, evenly spaced in $\ln
\lambda$, and (in short runs) measure the distribution of overlaps
$p(N_o|\lambda)$ for each. From this set we then pick out those values
of $\lambda$ for which successive $p(N_o|\lambda)$ exhibit an overlap by
area of approximately $20\%$. This criteria yields a suitable set of
intermediate states. 

Efficiency benefits result from noting that the rate of transitions in
$m$ depends on how quickly the number of overlaps $N_o$ relaxes after
each successive transition. To enhance this relaxation we preferentially
perform grand canonical insertions and deletions of small particles
within a spherical sub-volume of diameter $1.2\sigma_b$ centred on the
second big particle. Updates within the sub-volume occur with a
frequency $100$-fold that outside the sub-volume. Our approach --which
satisfies detailed balance-- greatly reduces the time spent updating
small particles whose coordinates are relatively unimportant for the
quantity we wish to estimate.


A further innovation, applicable to highly size-asymmetrical hard sphere
mixtures, stems from the observation that it is not actually necessary
to insert a big hard sphere in order to calculate the effective
potential. Instead it is sufficient and (generally much more efficient)
to  insert a hard {\em shell} of infinitesimal thickness. The basic idea
is that when fully inserted a hard shell particle encloses a number of
small particles. These remain in equilibrium with the reservoir (i.e.
their number can still fluctuate), but are fully screened from the rest
of the system because their surfaces cannot penetrate the shell wall.
Accordingly their contribution to the free energy is independent of the
position of the shell particle and hence their net effect is merely to
shift the value of the additive constant in the measurement of
$\mu_{ex}(r_{bb})$. Since the latter is anyway set by hand to ensure
that $\lim_{r_{bb}\to\infty}W(r_{bb})=0$, one does not need to know the
contribution to the free energy from the enclosed particles. 

From a computational standpoint, the task of inserting a hard shell is
much less challenging than that of inserting a hard sphere: essentially
the chemical potential grows with the particle size ratio like
$(1/q)^2$ rather than $(1/q)^3$.  Consequently, far fewer intermediate
stages $M$ are required to effect the insertion, which reduces
substantially the computational expenditure in measuring $\mu_{ex}(r)$
accurately. 

Whilst the staged insertion technique is not as straightforward to
implement as the GCA, it does not suffer the rapid decrease in
efficiency for $\eta_s^r > 0.2$ observed in highly asymmetrical
mixtures. It therefore allowed us to attain (for $q=0.1$), the
considerably larger reservoir packing fraction of $\eta_s^r= 0.35$.

\subsection{Correcting for finite size effects}
\label{sec:fseffect}

The effective potential $W(r)$ between two big particles is defined
in terms of the radial distribution function $g(r)\equiv g_{bb}(r)$,
with $r=r_{bb}$,  measured in the limit of infinite
dilution

\be
-\beta W(r)=\lim_{\rho_b\to 0}\ln[g(r)] \:,
\ee
for $r>\sigma_b$. In our simulation studies this limit is approximated by
placing a single pair of big hard spheres in the simulation box.  A
finite-size estimate to $g(r)$, which we shall denote $g_L(r)$, is then
obtained by fixing the first of these particles at the origin and
measuring (in the form of a histogram) the probability of finding the
second big particle in a shell of radius $r\to r+dr$, i.e.

\be
g_L(r)=\frac{P(r)}{P_{ig}(r)}\:,
\label{eq:g_r}
\ee
where, the normalization relates to the probability of finding an ideal gas
particle at this radius:

\be
P_{ig}(r)=\frac{4\pi r^2}{V}\:.
\label{eq:Pig}
\ee

Now the principal source of finite-size error in $g_L(r)$ arises from
the normalization of $P_{ig}$ by the system volume. Specifically, for a
finite sized system, the volume occupied by the hard sphere at the
origin is inaccessible to the second particle. Accordingly, the
accessible system volume is 

\be
\tilde{V}=V-v_1\:,
\label{eq:tV}
\ee
where $v_1=(1/6)\pi\sigma_b^3$ is the hard sphere volume. More
generally, one should define an {\em effective} excluded volume $\tilde
{v}_1$ for use in Eq.~\ref{eq:tV}, which allows
for the fact that the small particles can mediate additional repulsions and/or
attraction between the two big particles. In principle $\tilde{v}_1$
is given by

\be
\tilde{v}_1=4\pi\int_0^{\infty}[1-g(r)]r^2 dr\:.
\label{eq:v1eff}
\ee

It follows from Eqs.~\ref{eq:g_r}-\ref{eq:v1eff} that the principal
finite-size contribution to $g_L(r)$ is just an overall scale factor:

\be
g(r)=\frac{\tilde V}{V}g_L(r)\:.
\label{eq:grcorr}
\ee
Accordingly $g_L(r)$ approaches $V/\tilde{V}$ at
large $r$ instead of unity, whilst the calculated effective potential,
$W(r)$ decays to $\ln(\tilde{V}/V)$ instead of zero. 

One can conceive of a number of possible approaches for dealing with this
finite-size error. One is simply to minimize it by choosing a very large
system volume $V$ so that $V/\tilde{V}$ is close to unity. The
disadvantage of this approach is that in a size asymmetrical mixture, in
which the big particles are in equilibrium with a reservoir of small
ones, a huge number of small particles will necessarily fill the extra
space available in a bigger box. All the interactions arising from these
small particles then need to be computed -- which can become
prohibitively expensive. 

Another route, which we have adopted in the present work, is to attempt
to correct $g_L(r)$  by estimating the overall scale factor in
Eq.~\ref{eq:grcorr}, thus ensuring that $g(r)\to 1$ at large $r$. An
expedient approach to doing so, which utilizes as much as possible of
the information in $g_L(r)$, proceeds by determining the cumulative
integral of $g_L(r)$:

\be
G(R)=\int_0^{R}g_L(r)dr \:. 
\ee

\begin{figure}[h]
\includegraphics[type=pdf,ext=.pdf,read=.pdf,width=0.98\columnwidth,clip=true]{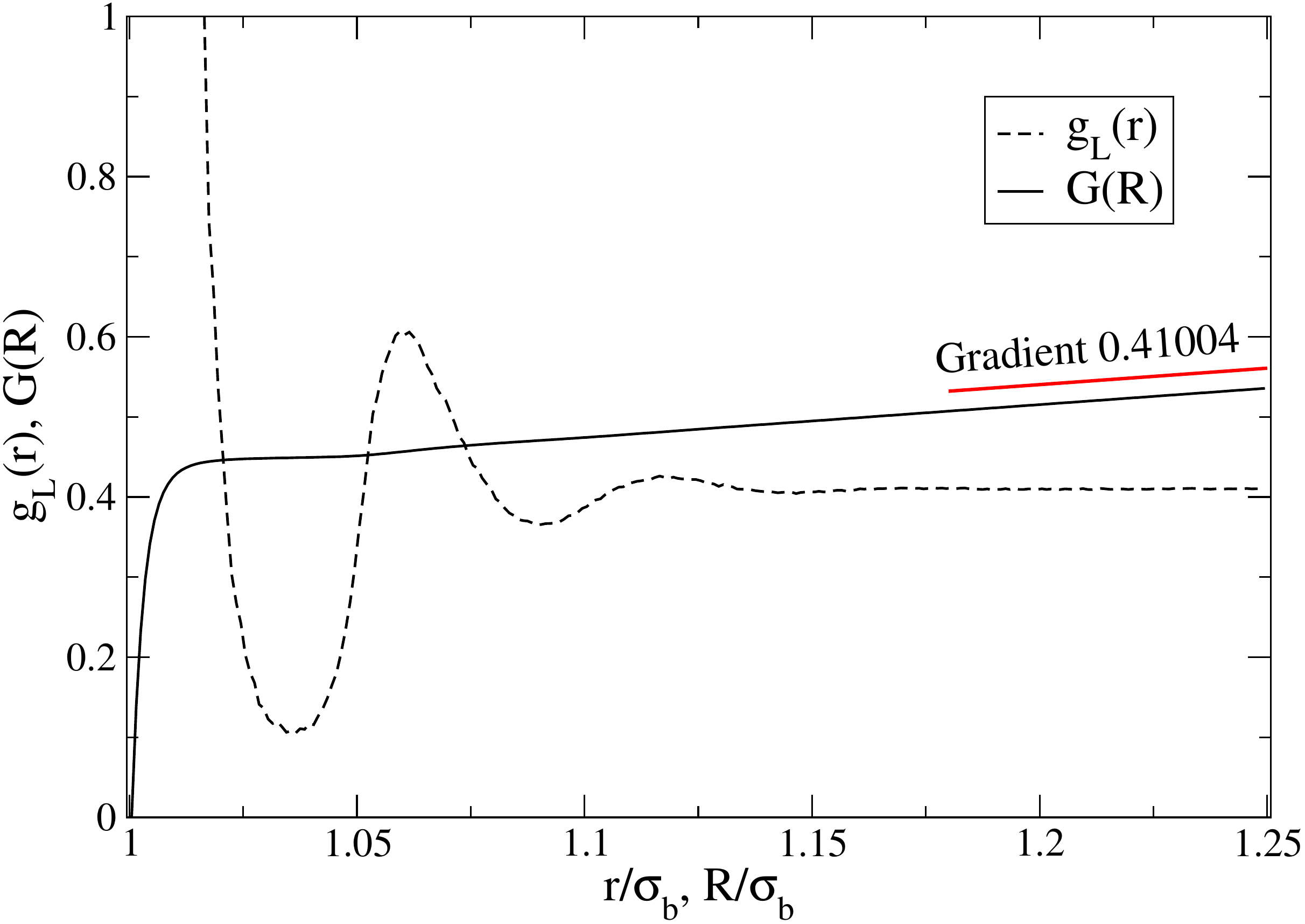}\\
\caption{The measured form of $g_L(r)$ for $q=0.05$,
$\eta_s^r=0.2$ obtained for a system size $L=2.5\sigma_b$. Also shown is the cumulative integral
$G(R)=\int_0^{R}g_L(r)dr$, together with an asymptotic linear fit, the gradient of
which yields the finite-size correction factor for $g(r)$.}
\label{fig:G_r} 
\end{figure}

In practice, this integral was observed to tend towards a smooth linear
form quite rapidly as the upper limit $R$ increases, a fact illustrated
for typical data in Fig.~\ref{fig:G_r}. A little thought then shows that
if when left uncorrected $g(r)$ tends to $V/\tilde{V}$ at large $r$, the
limiting gradient $m$ of $G(R)$ is $m=V/\tilde{V}$, which thus provides
the requisite correction factor for use in Eq.~\ref{eq:grcorr}. Thus one
corrects the measured histogram $g_L (r)$ by first fitting $G(R)$ to
obtain an estimate of the limiting gradient of the linear part, and then
scaling $g_L(r)$ according to $g(r)= m^{-1}g_L (r)$.

\section{Theoretical Methods}
\label{sec:theory}

As mentioned in the Introduction we choose to make comparisons between
our simulation results and those from the DFT insertion method and from
the morphometric and Derjaguin approximations.

\subsection{DFT}
\label{sec:DFT}

The DFT insertion method is described in detail in Roth {\em et al}
\cite{Roth2000a}. It is based on an exact result from the potential
distribution theorem, for an arbitrary mixture, that expresses the
effective potential between two big particles in terms of the one-body
direct correlation function of the big species in the limit where the
density $\rho_b$ of that species vanishes. For DFT treatments of hard
sphere mixtures that employ the fundamental measures theory (FMT)
approach \cite{RosenfeldPRL1989} the calculation of the effective
potential requires the computation of the density profile $\rho_s(r)$ of
the small spheres in the neighbourhood of a single fixed big sphere as
well as knowledge of the weight functions and the excess free energy
density of the (binary) mixture \cite{Roth2000a}. The FMT must be
sufficiently accurate to describe a very asymmetric binary mixture, i.e.
one with small values of $q$, in the limit $\rho_b\to 0$. In the
original DFT studies \cite{Goetzelmann1999, Roth2000a} the Rosenfeld
(RF) FMT \cite{RosenfeldPRL1989} was employed. In the present work we
employ RF, the White Bear (WB) version \cite{Roth:2002fk} and its
modification the White Bear Mark 2 (WB2) version
\cite{Hansen-Goos:2006uq}. These versions differ from RF in the choice
of the coefficients $\phi_\alpha$ entering the excess free energy
function. RF yields thermodynamic quantities that are the same as those
from Percus-Yevick (compressibility) approximation, whereas WB incorporates the accurate
Mansoori-Carnahan-Starling-Leland (MCSL) empirical bulk equation of
state. In WB2 additional self-consistency requirements are imposed on
the pressure. The consistency of the WB2 version was demonstrated in
calculations of the surface tension and other interfacial thermodynamic
coefficients for a one-component hard-sphere fluid adsorbed at a hard
spherical surface \cite{Hansen-Goos:2006uq}. Ref.~\cite{Roth:2010vn} provides an overview
of  recent developments and describes comparisons between different
versions of FMT.

Bo\c{t}an {\em et al}~\cite{Botan2009} carried out DFT insertion method
calculations as well as explicit (brute force) free energy minimization
for two fixed big spheres using different versions of FMT. These authors
provide a compendium of the ingredients entering the FMT functionals as
well as the thermodynamic coefficients required in the morphometric
approximation and we refer readers to Appendix B of \cite{Botan2009} for the explicit
formulae used in the present calculations. Their paper is important in
pointing to the regimes where the DFT insertion method is likely to
fail. In particular for $\eta_s^r = 0.419$ and $q=0.1$ and $0.2$ there are
substantial differences between the results for the depletion potential
calculated by brute force and from the insertion method. At higher
reservoir packing fractions the differences can be even larger. Moreover
different FMTs can give rise to quite different potentials at high small
sphere packings. The comparisons made by Oettel {\em et al}~\cite{Oettel2009} for the
depletion force using the RF functional suggest that  for $q=0.05$ the
insertion method is not especially accurate at $\eta_s^r =0.314$ and $0.367$. Of
course one is assuming that the brute force minimization is the more
accurate method as this requires only a reliable functional for a single
component hard sphere fluid-not one for the asymmetric mixture.
 
However, our present study focuses on smaller values of $\eta_s^r$ than
those considered in \cite{Botan2009}. Previous studies
\cite{Goetzelmann1999,Roth2000a} showed generally good agreement between
DFT insertion results and those of simulation for $q=0.1$ and $0.2$ and
$\eta_s^r$   typically up to $0.3$. Since we are concerned primarily
with investigating the depletion potential for highly asymmetrical
mixtures in regimes, accessible to simulation, near the onset of
fluid-fluid phase separation, we do not concern ourselves with very high
values of $\eta_s^r$ where the DFT insertion method is likely to be
inaccurate.

Another way of viewing this DFT insertion method is that it is
equivalent \cite{Goetzelmann1999,Roth2000a} to calculating the big-big
radial distribution function $g_{bb}(r)$ for a binary mixture using the
test particle route, i.e. one fixes a big sphere at the origin and
computes the inhomogeneous density profile of the big spheres
$\rho_b(r)$ by minimizing the mixture free energy functional for this
spherical geometry. Then $g_{bb}(r; \rho_b) = \rho_b(r)/ \rho_b$ and the
depletion potential is given by 

\be
-\beta W(r) = \lim_{\rho_b\to 0}    \ln g_{bb}(r; \rho_b) \: ,\hspace*{6mm} {\rm for~} r > \sigma_b .
\ee

In Refs.~\cite{Goetzelmann1999,Roth2000a}, for all cases considered, it was
demonstrated that a bulk packing fraction $\eta_b= 10^{-4}$ of the big
spheres was sufficiently small to ensure that the depletion potential
calculated from $g_{bb}(r)$ had converged to the limiting form. In a
very recent paper Feng and Chapman \cite{Feng2011} used the mixture WB
theory to calculate $g_{bb}(r)$ via the test particle route. For size
ratios $q=0.1$ they report good agreement with existing simulation
results \cite{Alawneh:2008kx} for concentrations of the big hard spheres
as small as $0.002$ and total packing fractions as large as $0.4$.
However, the packing fraction $\eta_b$ is still too high to be
appropriate for determining the depletion potential.

Roth {\em et al} \cite{Roth2000a} also introduced a parameterized form for the
depletion potential obtained from their DFT insertion method
calculations. Their motivation was to provide an explicit form
$W=1/2(1/q+1)\; \tilde{W}(x, \eta_s^r )$ , with $x=h/\sigma_s$ and
$h=r-\sigma_b$ the separation between the surfaces of the big spheres,
that would be valid for a range of size ratios $q$ and reservoir packing
fractions $\eta_s^r$ and therefore efficacious in studies of the phase
behaviour and correlation functions of binary hard sphere mixtures. The
authors were influenced by the comprehensive simulation studies of
Dijkstra {\em et al} \cite{Dijkstra1999} which had employed a very simplified
(third order virial expansion ) formula for the depletion potential
derived in ref.~\cite{Goetzelmann1998}. Roth {\em et al} aimed to provide a
formula, convenient for simulations of an effective one-component fluid,
that  captured both the short-ranged depletion attraction and the
long-ranged oscillatory behaviour of $W(r)$. Such a formula is of course
also useful in making comparisons between theory and experimental
measurements of the depletion potential. Their formula for $\tilde{W}$ consists
of a third order polynomial at small $x$ and an exponentially damped
oscillatory function at large $x$ accounting for the correct asymptotic
decay \cite{Roth2000a}. Comparisons made for $\eta_s^r$ between $0.1$ and
$0.3$ and different values of $q$ showed that the parametrized form gave
a good fit to the results of the numerical calculations. Largo and
Wilding \cite{Largo2006} employed this parametrized form in  simulation
studies of the (metastable) fluid-fluid critical point of the effective
one-component fluid, comparing their results with those from the much
simpler parametrized form used in \cite{Dijkstra1999}.

In the present study we noticed that the parametrization in
\cite{Roth2000a} did not recover the correct Asakura-Oosawa limiting
behaviour as $\eta_s^r\to 0$ and this restricts its regime of
application. Since we are concerned with making direct quantitative
comparisons with the results of our new simulations we performed new
numerical DFT insertion calculations, avoiding parametrization. 

\subsection{Derjaguin and Morphometric Approximations}
\label{sec:der_morph}

A much used theoretical ‘tool’ of colloid science is the Derjaguin
approximation \cite{Derjaguin1934} that relates the force between two
large convex bodies immersed in a fluid consisting of much smaller
particles or molecules to the integral of the excess pressure of the
same fluid contained between two parallel ‘walls’. In recent years
there has been considerable discussion about the regime of validity of
the Derjaguin approximation for our present case of a fluid of small
hard spheres confined between two fixed big hard spheres or between a
planar hard wall and a single big hard sphere. The reader is referred to
Refs.~\cite{Goetzelmann1998,Henderson2002,Roth2000a,Herring07,Oettel2004} should
they wish to savour the arguments. Herring and Henderson
\cite{Herring06,Herring07} performed simulations for the wall-sphere
case for $q=0.05$ and $\eta_s^r = 0.3$ and $0.4$, comparing their
results for the depletion force with those from the Derjaguin
approximation and from the DFT insertion method \cite{Roth2000a}. In the present
work we perform equivalent comparisons, for the sphere-sphere case,
using what we believe is much more accurate simulation data for the
depletion potential.

As shown in \cite{Goetzelmann1998} the depletion potential {\em
difference} for hard spheres obtained from the Derjaguin approximation
can be expressed succinctly as:

\begin{widetext}
\be
W_{Der}(h) -W_{Der}(\sigma_s) = -\frac{\epsilon\pi}{2} (\sigma_s + \sigma_b)(\sigma_s
-h)\left[\frac{1}{2} p(\eta_s^r)(\sigma_s-h) +2\gamma(\eta_s^r) \right];\hspace*{3mm} 0<h<\sigma_s
\label{eq:der_dep}
\ee
\end{widetext}
where $h$ is the separation between the surfaces, $p(\eta_s^r)$ is the pressure
of the small sphere reservoir and $\gamma(\eta_s^r)$ is the surface tension between a
single planar hard wall and the small sphere fluid. Within the Derjaguin
approximation the potential between a wall and a single big sphere is
precisely twice that between two big spheres: $\epsilon$ is 1 for sphere-sphere
and $2$ for wall-sphere. Expressions for the pressure and surface tension
are listed in Appendix A of \cite{Botan2009}. Another expression for the surface
tension due to D. Henderson and Plischke \cite{Henderson08041987} as obtained by fitting
simulation data was used in \cite{Herring07}. For $h>\sigma_s$  the depletion potential
depends on the excess grand potential of the small sphere fluid confined
in the planar hard wall slit which must be obtained from simulation or
DFT \cite{Botan2009}. 

Morphometric thermodynamics \cite{Konig:2004uq} was developed to calculate the
solvation free energy (excess grand potential) of large convex bodies
immersed in a solvent. Its application to determining depletion
potentials is described in \cite{Oettel2009,Botan2009} where it is shown that 

\be 
W_{Morp}(h) = -p\Delta V(h)- \gamma \Delta A(h)-\kappa \Delta C(h)-4\pi\bar{\kappa},
\label{eq:fullmorph}
\ee 
for $0<h< \sigma_s$. Here $\Delta V(h)$ and
$\Delta A(h)$ are the volume and surface area of the overlap of
exclusion (depletion) zones around the big spheres (or a wall and a big
sphere) and $\Delta C(h)$ is the integrated mean curvature of the
overlap volume. The thermodynamic coefficients are the pressure $p$,
surface tension $\gamma$ and the two bending rigidities $\kappa$ and 
$\bar{\kappa}$ ; these four quantities are functions of $\eta_s^r$ .The
fourth term is the difference in integrated Gaussian curvatures between
a dumbbell ($4\pi$) and two disconnected spheres ($8\pi$). For $h>
\sigma_s$ the dumbbell separates into two disconnected spheres. Thus
$W_{Morp}(h> \sigma_s) = 0$. Explicit formulae are given in
Ref.~\cite{Botan2009} for the geometrical quantities and for the four
thermodynamic coefficients. Note that
$W_{Morp}(\sigma_s^-)=-4\pi\bar{\kappa}(\eta_s^r)$, independent of size
ratio. This term is  small in comparison with the others.

In order to connect with the Derjaguin approximation we invoke the
colloidal limit , i.e. $q\to 0$. Then the difference

\begin{widetext}
\be
W_{Morp}(h)-W_{Morp}(\sigma_s) = -\frac{\epsilon\pi}{2} (\sigma_s + \sigma_b)(\sigma_s -h)\left[\frac{1}{2} p(\eta_s^r)(\sigma_s-h)+2\gamma(\eta_s^r)\right]-\kappa(\eta_s^r)\pi^2\sqrt{\epsilon(\sigma_s-h)(\sigma_s+\sigma_b)/2}, 
\label{eq:Der_Morph}
\ee
\end{widetext}
for $0<h <\sigma_s$. 
Since $\kappa$ is positive the morphometric approach contributes an
additional attractive term, augmenting the attraction from the pressure
($\Delta V(h)$) term. $\gamma$ is negative so the surface tension
($\Delta A(h)$) term gives a repulsive contribution to the depletion
potential. The physical interpretation of the third term in
Eq.~\ref{eq:fullmorph} or ~\ref{eq:Der_Morph} is of a line contribution to the effective
interaction associated with the circumference of the edge of the annular
wedge formed between the two exclusion spheres where the line tension is
$-\kappa\pi/2$ \cite{Oettel2009}. As this term is proportional to
$\sqrt{\epsilon}$ (not to $\epsilon$ ) Derjaguin scaling is violated
\cite{Botan2009}. 

The morphometric analysis must break down in the limit $h\to\sigma_s$
where Eq.~\ref{eq:fullmorph} or ~\ref{eq:Der_Morph} predicts that the depletion force is singular,
diverging as ($\sigma_s -h)^{-1/2}$. The reasons for this unphysical
limiting behaviour are associated with problems of self-overlapping
surfaces as explained in Refs.~\cite{Botan2009,Oettel2009}. However,
away from this limit one might expect the elegant geometrical  arguments
underlying the morphometric analysis to capture the essential physics.
Indeed the comparisons with brute force DFT results for the depletion
force in Ref.~\cite{Oettel2009} indicated rather good agreement for a
range of $q$ and $\eta_s^r = 0.314$.

In Sec.~\ref{sec:results}  we compare the results of 
Eqs.~\ref{eq:der_dep} and \ref{eq:fullmorph} with our simulation data
and with results from the DFT insertion method.

\section{Results}
\label{sec:results}
\subsection{Test Case}

We have tested the ability of the GCA to accurately determine effective
potentials by applying it to the case of the Asakura-Oosawa (AO) Model
\cite{Asakura1954,Asakura:1958uq}. This model describes colloidal
hard-spheres in a solvent of non-interacting point particles modelling
ideal polymer that have a
hard-particle interaction with the colloids. Although not the case of
additive hard-spheres which is our primary focus in this paper, the
extremely non-additive AO model does provide a very useful testbed for our simulation methodology
because the exact form of the depletion potential is known, taking the
form: \cite{Asakura:1958uq}

\begin{widetext}
\be
\beta W_{\rm AO}(r)=\left \{ \begin{array}{ll} 
-\eta_s^r\frac{(1+q)^3}{q^3}\left[1-\frac{3r}{2\sigma_b(1+q)}+\frac{r^3}{2\sigma_b^3(1+q)^3}\right]\mbox{\hspace{1mm}},   &  \sigma_b< r< \sigma_b+\sigma_s \\
                               &       \\
 0, & r\ge  \sigma_b+\sigma_s \, ,\\
\end{array}
\right.
\label{eq:AOpot}
\ee
\end{widetext}
where $\sigma_s$ is the `polymer' diameter, i.e. the colloid-polymer
pair potential is infinite for $r<(\sigma_b+\sigma_s)/2$.

\begin{figure}[h]
\includegraphics[type=pdf,ext=.pdf,read=.pdf,width=1.0\columnwidth,clip=true]{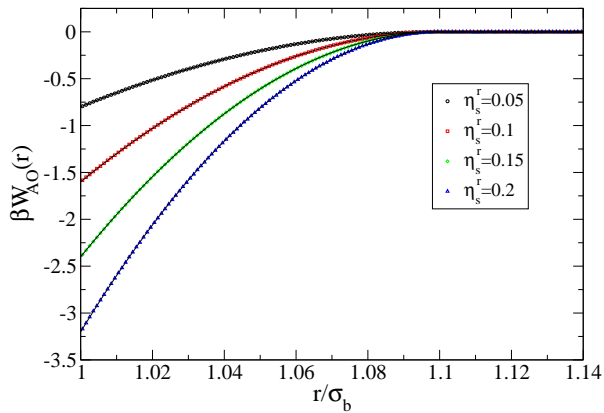}\\

\caption{A comparison of GCA simulation measurements of the depletion
potential of the AO model with the exact analytical form. The
size ratio is $q=0.1$ and data are shown for four values of the
reservoir packing fraction. Symbols are results from GCA measurements of
$g(r)$ for a pair of big particles, corrected for finite-size effects
and transformed via $\beta W(r)=-\ln[g(r)]$. Statistical errors are
smaller than the symbol size. Lines are the exact AO effective
potential, Eq.~\ref{eq:AOpot}.}

\label{fig:AOres} 
\end{figure}

Simulation measurements of $g(r)\equiv g_{bb}(r)$ were performed for the
AO model using the GCA for a system comprising a pair of hard spheres in
a cubic box of linear dimension $L=3\sigma$ in equilibrium
with a reservoir of small particles having size ratio $q=0.1$. Since the
small particles are mutually non-interacting, the chemical potential of
the reservoir is just that of an ideal gas. The depletion potential was
calculated as $\beta W(r)=-\ln[g(r)]$ and the  results were corrected
for finite-size effects according to the procedure described in
Sec.~\ref{sec:fseffect}. In Fig.~\ref{fig:AOres} we compare the results
of simulations of the effective potential with the exact result. Data is
shown for various values of the reservoir packing fraction. In each
instance, the simulation results (symbols) are indistinguishable from
the analytical form (lines) within the very small statistical errors, a
finding that supports the validity and accuracy of the simulations.

\subsection{Effective potentials for additive hard spheres}

We turn now to our measurements of the effective potential for highly
size asymmetrical additive hard spheres and the comparison with DFT
calculations. Similarly to the case of the AO model, our simulations
treat the small particles grand canonically, i.e. their number
fluctuates under the control of a chemical potential $\mu_s^r$. The
value of $\mu_s^r$ is chosen to yield some nominated value of the packing
fraction of small particles, $\eta_s^r$, in the notional reservoir. Thus
the simulations require prior knowledge of $\mu_s^r(\eta_s^r)$. In
principle, one could employ the Carnahan-Starling (CS) approximation
\cite{Hansen-MacDonald} to estimate the requisite chemical potential. 
However in tests we found this approximation to be insufficiently
accurate for our purposes. For instance, taking $\eta_s^r = 0.32$ as an
example, if we employ the CS value for the chemical potential we
actually measure $\bar{\eta}_s^r = 0.3195$, which while close to the
target, lies outside the range of fluctuations in $\eta_s^r$ that occur
in a large simulation box. In order to determine $\mu_s^r$ more
accurately we therefore performed a series of accurate grand canonical
simulations for the pure fluid of small hard spheres in a large box of
$L=50\sigma_s$. We then employed histogram reweighting to extrapolate to
the precise values of the chemical potential that corresponds to the
various values of $\eta_s^r$ that we wished to study. These resulting
estimates are listed in table~\ref{tab:eos}.

\begin{table}[h]
\begin{tabular}{c|c}
\hline
$\eta_s^r$ & $\beta\mu_s^r$\\\hline
 0.05 & -1.9079(1)\\
 0.10 & -0.6770(3)\\
 0.15 & 0.3923(2)\\
 0.20 & 1.5105(1)\\
 0.32 & 5.0472(1)\\
 0.35 & 6.2659(2)\\
 \hline
\end{tabular}
\caption{Measured values of the reduced chemical potential
$\beta\mu_s^r$ corresponding to each of the packing fractions $\eta_s^r$
listed. The data were obtained by histogram reweighting the results of grand
canonical simulations of hard spheres obtained at the nearby value of
$\beta\mu_s^r$ predicted by the CS approximation. The simulation cell size was
$L=50\sigma_s$. The definition of $\mu_s^r$ is subject to the convention of   
choosing the thermal wavelength to equal the hard sphere
diameter.}
\label{tab:eos}
\end{table}

Measurements of the radial distribution function $g(r)$ were made for a
pair of big hard spheres in equilibrium with a reservoir of small hard
spheres, for the combinations of values of $\eta_s^r$ and size ratio $q$
shown in table~\ref{tab:combin}. The system size was $L=3\sigma_{b}$ for
$q=0.1$, while for  $ q=0.05,0.02,0.01$, where the range of the
depletion potential is shorter, $L=2.5\sigma_{b}$ was used. In all
cases the depletion potential was obtained as $\beta W(r)=-\ln[g(r)]$
with corrections for finite size effects applied as described in
Sec.~\ref{sec:fseffect}.

\begin{table}[h]
\begin{tabular}{c|cccccc}
\hline
\multicolumn{1}{c|}{$q$} & \multicolumn{6}{c}{$\eta_s^r$}\\\hline
0.10 & 0.05 & 0.1 & 0.15 & 0.20  & {\bf 0.32} & {\bf 0.35}\\
0.05 & 0.05 & 0.1 & 0.15 & 0.20  &  -         & -  \\
0.02 & 0.05 & 0.1 & 0.15 &  -    &  -         & -  \\
0.01 & 0.05 & - & -    &  -    &  -         & -  \\\hline
\end{tabular}

\caption{Combinations of particle size ratio $q$ and reservoir volume
fraction $\eta_s^r$ for which we compare simulation estimates of
depletion potentials with DFT predictions. Values shown in normal typeface were studied by
simulation using the GCA described in Sec.~\protect\ref{sec:gca}, while
those in boldface were studied using  staged insertion MC as described
in Sec.~\protect\ref{sec:staged}.}

\label{tab:combin}
\end{table}

\subsubsection{Comparison of simulation and density functional theory results}

\begin{figure}[h]
\includegraphics[type=pdf,ext=.pdf,read=.pdf,width=0.98\columnwidth,clip=true]{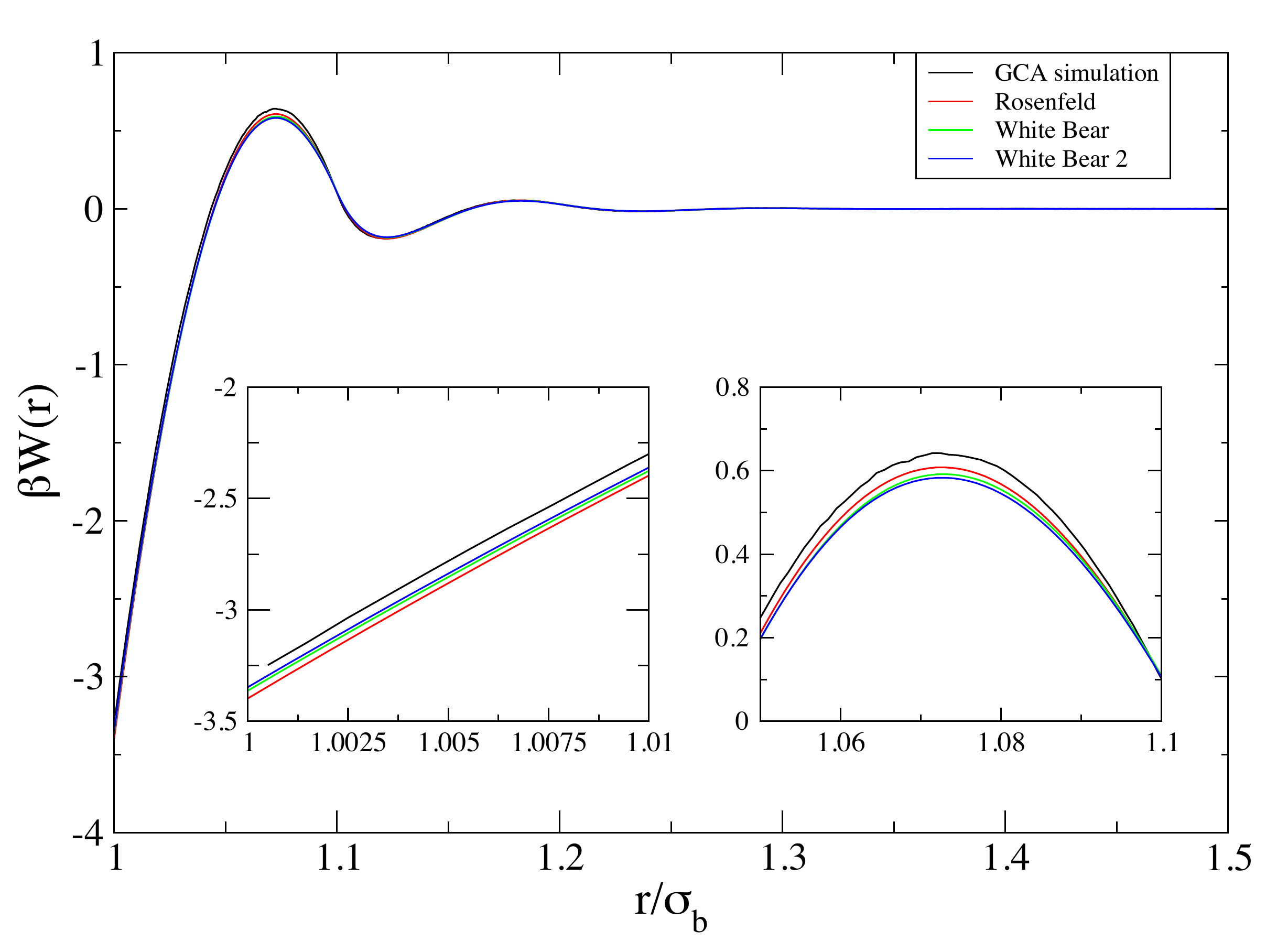}\\

\caption{Simulation and DFT results for the hard sphere depletion
potential $\beta W(r)$ for $q=0.1, \eta_s^r=0.2$. The abscissa is
the separation of hard sphere centres expressed in
units of the big particle diameter $\sigma_{b}$. The two insets expand
the region close to hard sphere contact (left panel) and around the first
maximum (right panel).}
\label{fig:q0.1-eta0.2} 
\end{figure}

We now examine a selection of the measured  effective
potentials.  Data for $q=0.1, \eta_s^r=0.2$ is shown in
Fig.~\ref{fig:q0.1-eta0.2}. Despite our use of a rather small histogram
bin size of just $\delta r=0.001$ to accumulate estimates of $\beta
W(r)$, the statistical fluctuation is sufficiently small that one can simply
connect the data points by lines. This allows us to better discern
differences between the simulation results and those of the DFT
calculations using the insertion method, which are also included on the plot. Data for three
versions of DFT are shown, namely the Rosenfeld (RF), White Bear (WB)
and White Bear 2 (WB2) functionals. Clearly for these parameters the
overall agreement is very good. To quantify the extent of the accord,
the two insets to Fig.~\ref{fig:q0.1-eta0.2} show a comparison in the
range of separations close to hard sphere contact (left inset) and
around the first maximum (right inset). These show that near contact, 
WB2, fares slightly better than WB, which is in turn better than RF.
Near the first maximum in the potential however, the trend is reversed
and RF has the greatest accord with the simulation data, while WB is
better than WB2. 

\begin{figure}[h]
\includegraphics[type=pdf,ext=.pdf,read=.pdf,width=0.98\columnwidth,clip=true]{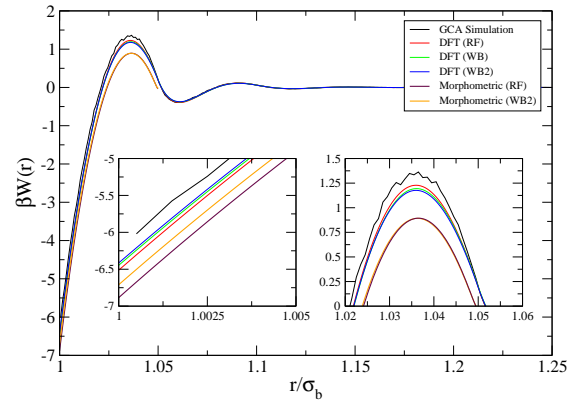}\\
\caption{As for Fig.~\protect\ref{fig:q0.1-eta0.2} but with $q=0.05,
\eta_s^r=0.2$. Also shown are the results of the morphometric
approximation Eq.~\ref{eq:fullmorph}.}
\label{fig:q0.05-eta0.2} 
\end{figure}

A similar picture emerges for $q=0.05, \eta_s^r=0.2$ as shown in
Fig.~\ref{fig:q0.05-eta0.2}. Although here the simulation data is not as
smooth as for $q=0.1$, the form and magnitude of the deviations from
the DFT are similar. We comment later on the results of the morphometric
approximation Eq.~\ref{eq:fullmorph}. 

\begin{figure}[h]
\includegraphics[type=pdf,ext=.pdf,read=.pdf,width=0.98\columnwidth,clip=true]{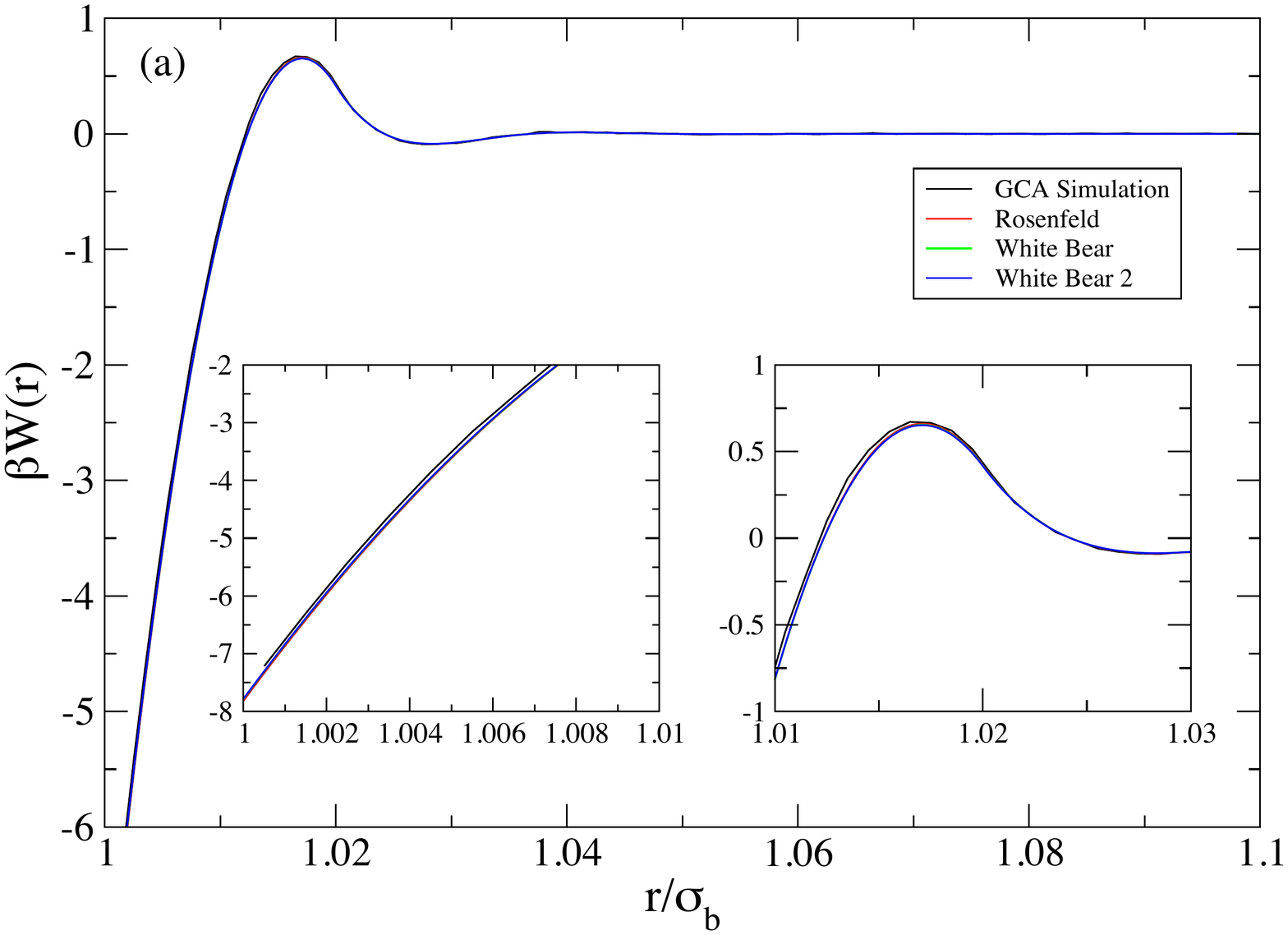}\\
\includegraphics[type=pdf,ext=.pdf,read=.pdf,width=0.98\columnwidth,clip=true]{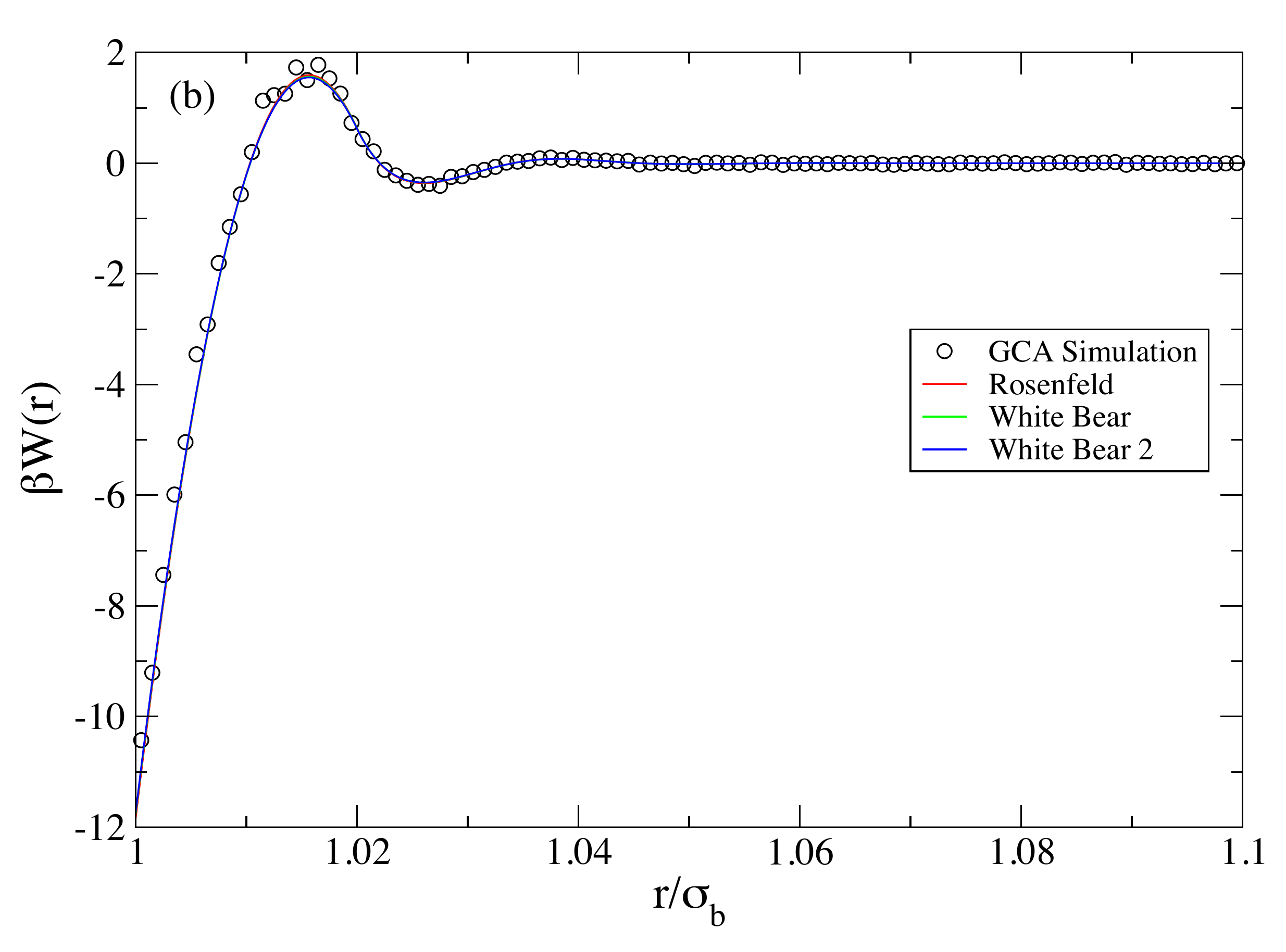}\\
\caption{As for Fig.~\protect\ref{fig:q0.1-eta0.2} but with {\bf (a)} $q=0.02, \eta_s^r=0.1$ and
{\bf (b)} $q=0.02,\eta_s^r=0.15$.}
\label{fig:q0.02} 
\end{figure}

\begin{figure}[h]
\includegraphics[type=pdf,ext=.pdf,read=.pdf,width=0.98\columnwidth,clip=true]{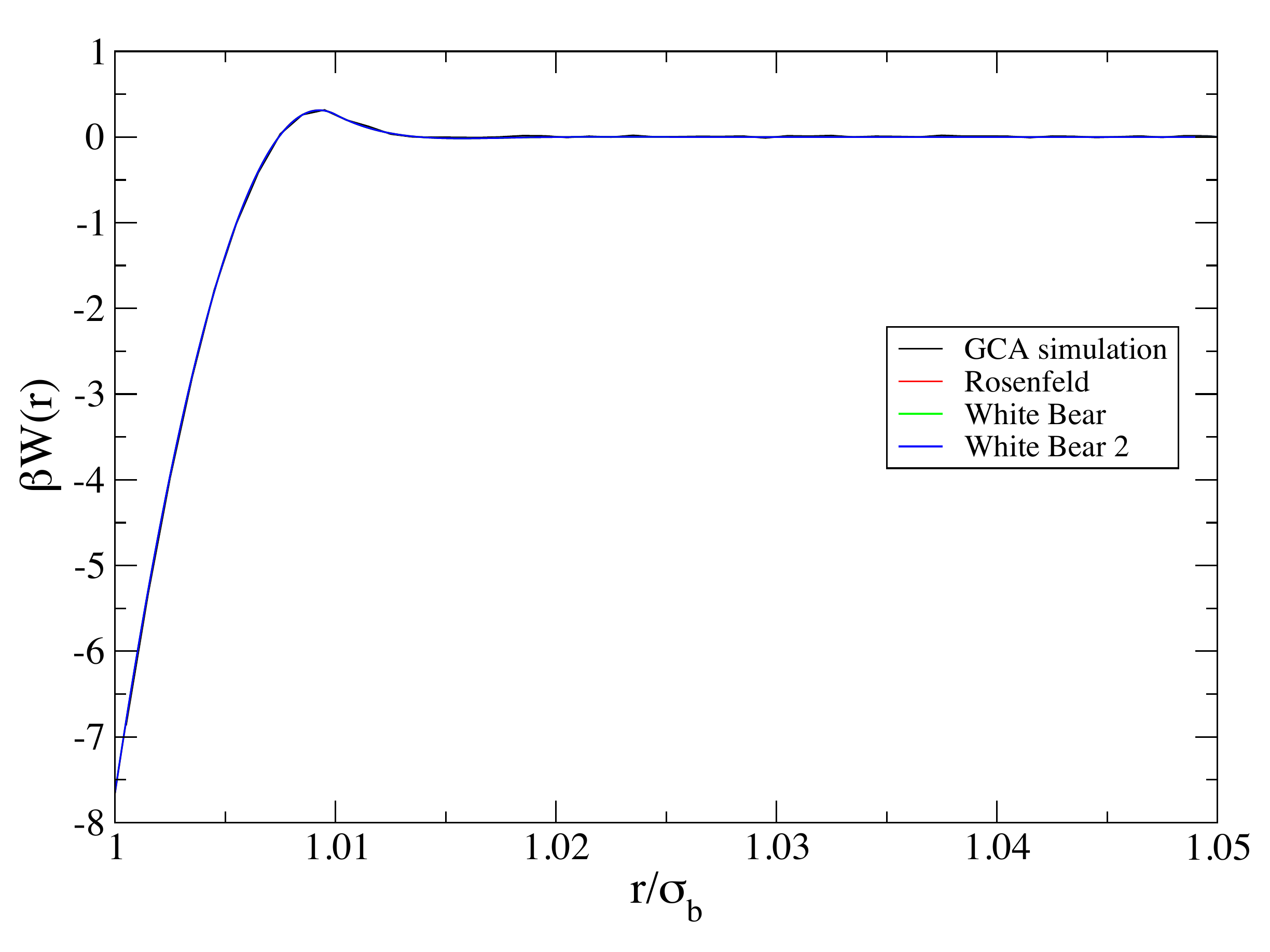}\\
\caption{As for Fig.~\protect\ref{fig:q0.1-eta0.2} but with $q=0.01, \eta_s^r=0.05$.}
\label{fig:q0.01} 
\end{figure}

Generally speaking, the smaller the size ratio, $q$, the lower the
maximum packing fraction $\eta_s^r$ for which we can obtain good
statistics with the GCA. Data for $q=0.02$, with $\eta_s^r=0.1$ and
$\eta_s=0.15$ is shown in Fig.~\ref{fig:q0.02}(a) and
~\ref{fig:q0.02}(b), respectively. For this size ratio and these (low)
small sphere packings the various versions of FMT perform very well.
Data for  $q=0.01$ with $\eta_s^r=0.05$ is shown in
Fig.~\ref{fig:q0.01}. In this extreme case the insertion DFT results are
almost indistinguishable from each other and from the results of
simulation. However one should note that there is still a maximum in
$W(r)$ -- one is not yet in the AO limit although the contact value is
close to the AO value given by Eq.~\ref{eq:AOpot}.

\begin{figure}[h]
\includegraphics[type=pdf,ext=.pdf,read=.pdf,width=0.98\columnwidth,clip=true]{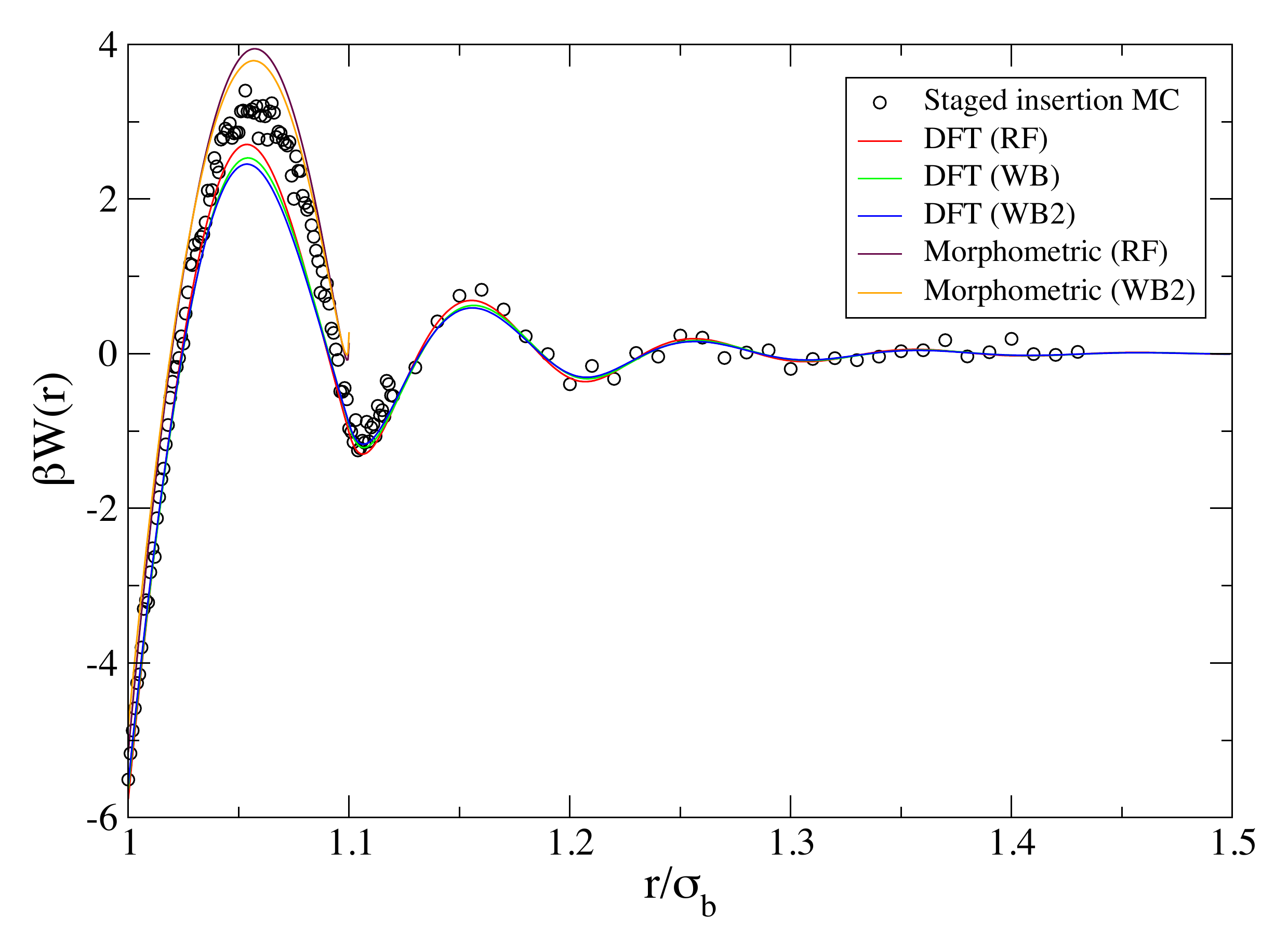}\\
\caption{The depletion potential for $q=0.1, \eta_s^r=0.35$
obtained using the staged insertion simulation method and DFT. The simulation data points
represent the results of $150$ independent measurements of $\beta W(r)$ made at
various fixed values of the big particle separation, though concentrated
in the range $r<1.12\sigma_b$. Also shown are the results of the morphometric
approximation Eq.~\ref{eq:fullmorph}.}
\label{fig:q0.1-eta0.35} 
\end{figure}

For our system, the GCA is operable for $\eta_s^r\lesssim 0.2$. To go
beyond this limit we have employed the staged insertion algorithm
outlined in Sec.~\ref{sec:staged}. Simulation results for $q=0.1, 
\eta_s^r=0.35$ are compared with those from DFT calculations in
Fig.~\ref{fig:q0.1-eta0.35}. While the simulation data are somewhat
noisier, they show that in this regime, quite significant discrepancies
with the DFT insertion method have emerged. The principal form of the discrepancy, i.e. DFT
underestimates the height of the first maximum, is similar in form but
greater in magnitude to that seen using the GCA at smaller values of
$\eta_s^r$ (cf. Fig.~\ref{fig:q0.1-eta0.2}). Once again RF fares better
than the two WB functionals but underestimates the first maximum by
about $0.5k_BT$. Results from the three functionals agree quite well with one
another and with simulation near contact.

\subsubsection{Comparison with Derjaguin and morphometric approximations}

In this subsection we make comparisons between our simulation and DFT
results with those from the approximations described in
Sec.~\ref{sec:der_morph}. Recall that the Derjaguin approximation is
specifically designed to tackle small size ratios. In
Fig.~\ref{fig:q0.05-eta0.2} we compare the results of the morphometric
approximation Eq.~\ref{eq:fullmorph} with those from simulation and DFT.
Two sets of thermodynamic coefficient were used: RF and WB2
\cite{Botan2009}. Both versions underestimate the maximum of the
depletion potential and overestimate the magnitude of the potential at
contact for $q=0.05$ and $\eta_s^r=0.2$. By contrast for $q=0.1$ and
$\eta_s^r=0.35$, Fig.~\ref{fig:q0.1-eta0.35} shows that both versions of
the morphometric approximation overestimate the maximum and underestimate
the magnitude of the potential at contact.
Fig.~\ref{fig:q0.1-eta0.35} also shows a pronounced minimum for $h$
close to $\sigma_s$. This feature is absent in both simulation and DFT.
It is associated with the unphysical divergence of the line tension
contribution to the depletion force arising in the morphometric
treatment. Recall that $W_{Morph}$ is zero for separations $h>\sigma_s$.

\begin{figure}[h]
\includegraphics[type=pdf,ext=.pdf,read=.pdf,width=0.95\columnwidth,clip=true]{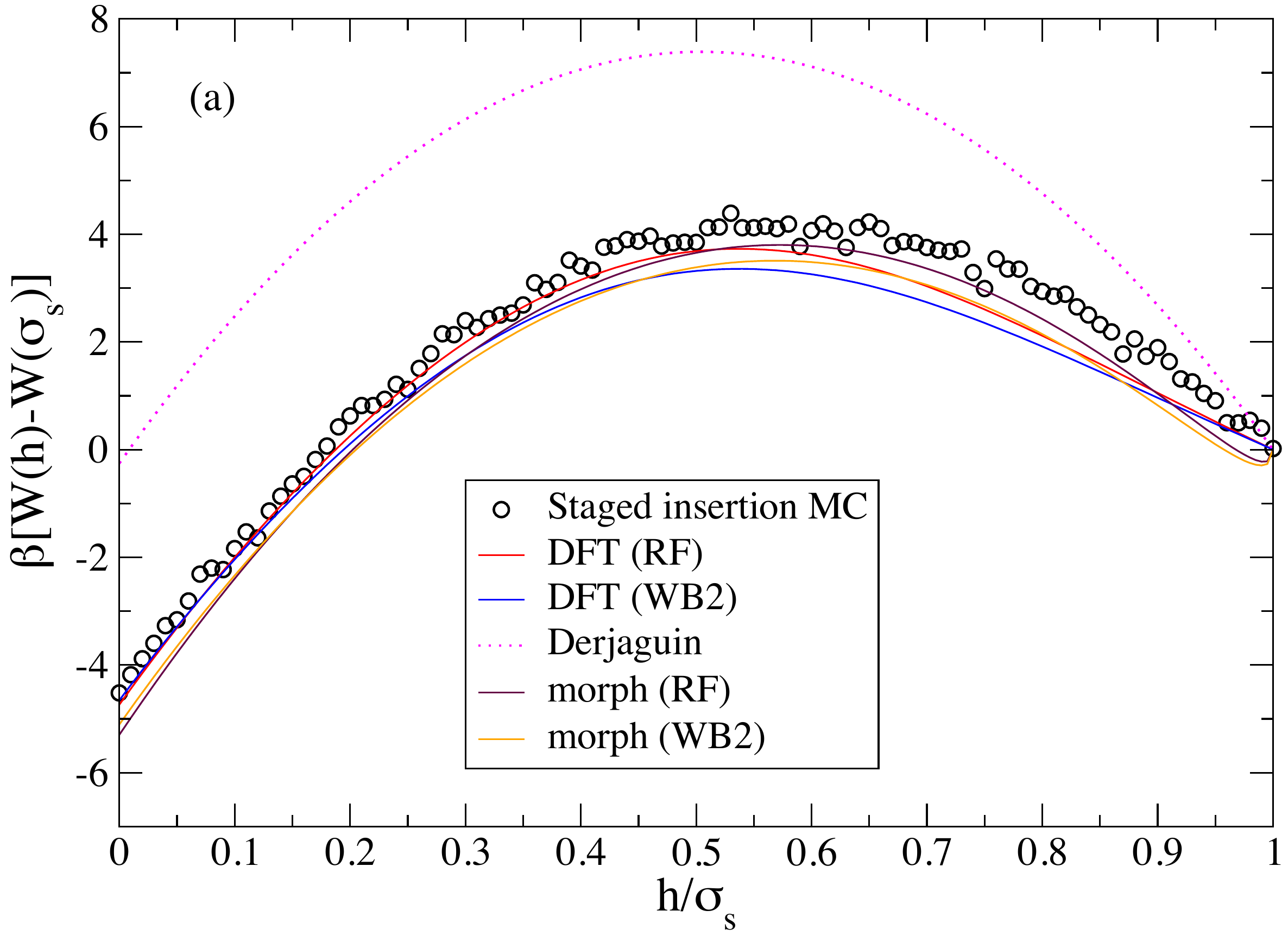}\\
\includegraphics[type=pdf,ext=.pdf,read=.pdf,width=0.95\columnwidth,clip=true]{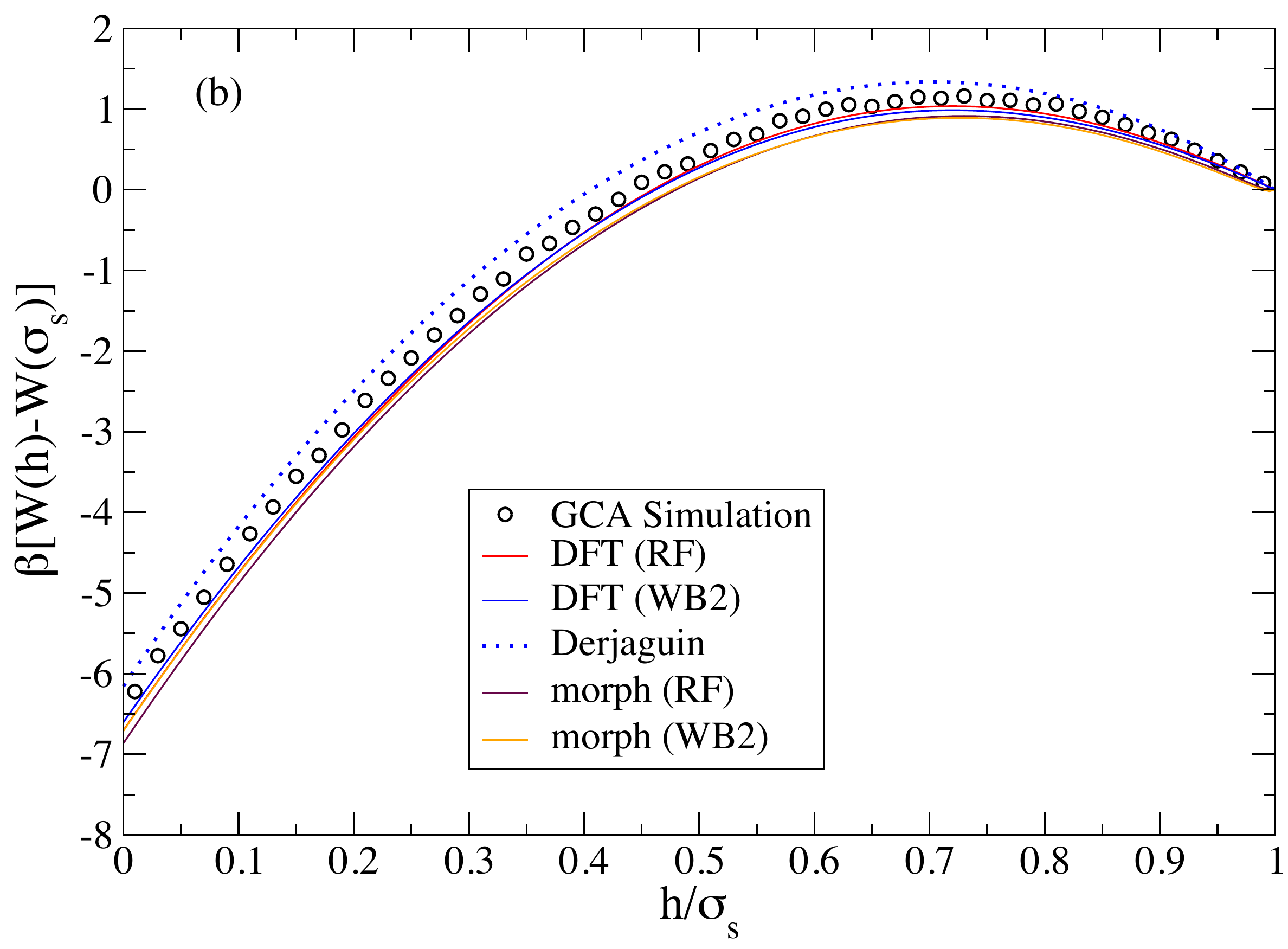}\\
\caption{Comparison of $\beta W(h)-\beta W(\sigma_s)$ obtained from
simulation and DFT for {\bf (a)} $q=0.1, \eta_s^r=0.35$ (cf
Fig.~\protect\ref{fig:q0.1-eta0.35}) and {\bf (b)} $q=0.05, \eta_s^r=0.2$ (cf
Fig.~\protect\ref{fig:q0.05-eta0.2}) with results of the Derjaguin
(Eq.~\ref{eq:der_dep}) and morphometric approximations (Eq.~\ref{eq:fullmorph}). }
\label{fig:derjag}
\end{figure}

Fig.~\ref{fig:derjag}(a) compares the depletion potential {\em
difference} obtained from simulation and insertion method DFT for
$q=0.1$ and $\eta_s^r=0.35$ with results from the Derjaguin
approximation (where plotting the difference is the natural choice
\cite{Goetzelmann1998}) and morphometric approximations,
Eqs.~\ref{eq:der_dep} and \ref{eq:fullmorph} respectively. For this
value of $q$ the packing fraction of the small spheres is sufficiently
large to enter the regime where fluid-fluid phase separation might
occur-as discussed below in Sec.~\ref{sec:B2}. Thus it is interesting to observe how well
these explicit approximations perform. Similar remarks apply for
$q=0.05$ and $\eta_s^r=0.2$ for which comparisons are presented in
Fig.~\ref{fig:derjag}(b)

One sees in Fig.~\ref{fig:derjag}(a) that the Derjaguin approximation is
very poor. Overall the morphometric approximations fare considerably
better than Derjaguin with RF better than WB2 near the maximum. However,
both morphometric versions overestimate the magnitude of the contact
value by about $0.5k_BT$. Note once again the minimum close to
$h/\sigma_s=1$ for this packing fraction. The situation is clearly
different in Fig.~\ref{fig:derjag}(b) where the Derjaguin and
morphometric approximations are reasonably good; they bracket the
simulation and DFT results. The two morphometric versions yield results
that are very close and even in this difference plot one sees that these
fall below the simulation results both at maximum and at contact. At
this smaller value of $\eta_s^r$ there is no minimum visible in the
depletion potential. Although plotting the difference appears to improve
the level of agreement between morphometric and simulation, one should
recall that it is the actual depletion potential displayed in
Figs.~\ref{fig:q0.05-eta0.2} and \ref{fig:q0.1-eta0.35} which matters,
eg. in determining $B_2(\eta_s^r)$, to which we now turn.

\subsubsection{Second virial coefficients}
\label{sec:B2}
While the various simulation and DFT estimates of effective potentials
show generally good agreement at low $\eta_s^r$, the differences grow
with increasing $\eta_s^r$ and it is natural to enquire as to the likely
implications for the properties of the bulk mixture and
in particular phase behaviour. A useful indicator in this regard is the
value of the second virial coefficient $B_2$:

\be
B_2=-2 \pi \int^{\infty}_{0} \left(e^{-\beta\phi_{\mathit{eff}}(r)}-1  \right) r^2 dr\:.
\label{eq:B2}
\ee
where the effective pair potential is defined in Eq.~\ref{eq:effpot}.

Previous work by Vliegenthart and Lekkerkerker \cite{Vliegenthart2000}
and Noro and Frenkel \cite{Noro2000} has shown that an extended
corresponding states behaviour applies to fluids that share the same
value of $B_2$. Specifically, the measured values of $B_2$ at 
fluid-fluid criticality were found to be similar across a wide range of
model potentials. Subsequent work by Largo and Wilding \cite{Largo2006},
examined this criterion explicitly for the case of two DFT-based hard
sphere effective potentials which had been fitted to analytical forms
and parameterized in terms of the reservoir packing fraction $\eta_s^r$ 
 \cite{Dijkstra1999,Roth2000a}. Using simulation of a single component
fluid interacting via a pair potential (Eq.~\ref{eq:effpot}) with $W(r)$
given by these parameterized depletion potentials, the value of
$\eta_s^r$ at which the metastable fluid-fluid critical point occurs was
determined using an accurate approach based on finite-size scaling
\cite{Largo2006,Wilding1995}. Interestingly for both $q=0.1$ and $0.05$
and both choices of parameterized potentials, the value of $B_2$ for the
depletion potential at criticality was in quantitative agreement with
that of the adhesive hard sphere model (AHS) at its fluid-fluid critical
point as determined separately in simulations by Miller and Frenkel
\cite{Miller2004}. These authors report a critical value
$B_2^{AHS}=-1.207B_2^{HS}$, where the hard spheres second virial
coefficient $B_2^{HS}=2\pi\sigma_b^3/3$. The level of agreement was much
greater than that seen for more general model potentials (such as square
well or Lennard-Jones model studied by Noro and Frenkel), suggesting
that the quasi-universality of the critical point $B_2$ value holds
particularly closely for effective potentials whose attractive piece is
very short ranged in nature, as pertains to highly size asymmetrical
hard sphere mixtures. Further confirmation of this has been found very
recently in simulations of the AO potential where, for $q=0.1$, Ashton
\cite{Ashton_unpub} has found that the metastable critical point occurs
at $\eta_s^r=0.249(1)$, to be compared with the prediction
$\eta_s^r=0.2482$ based on matching to $B_2^{AHS}$.

\begin{figure}[t]
\includegraphics[type=pdf,ext=.pdf,read=.pdf,width=0.85\columnwidth,clip=true]{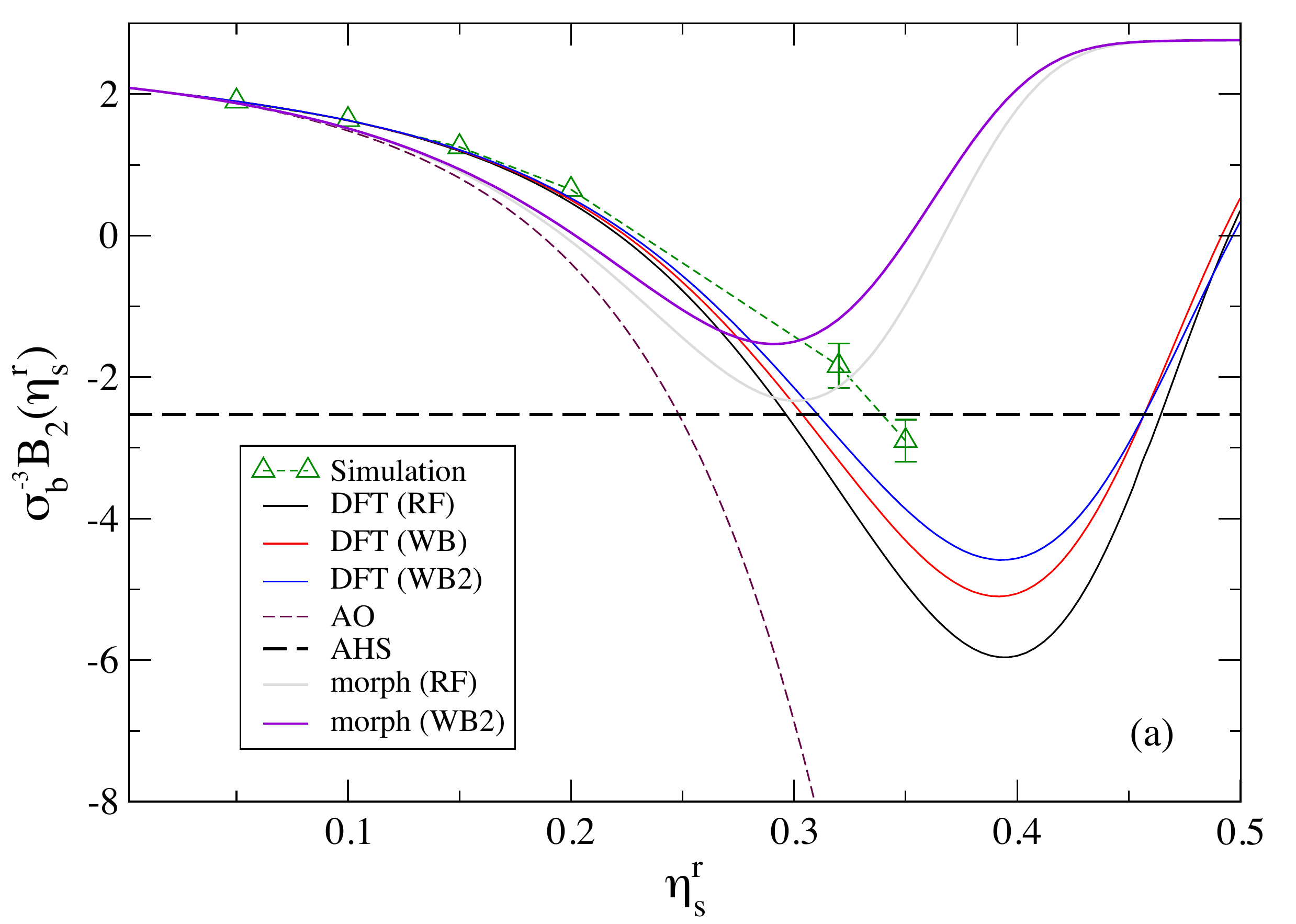}\\
\includegraphics[type=pdf,ext=.pdf,read=.pdf,width=0.85\columnwidth,clip=true]{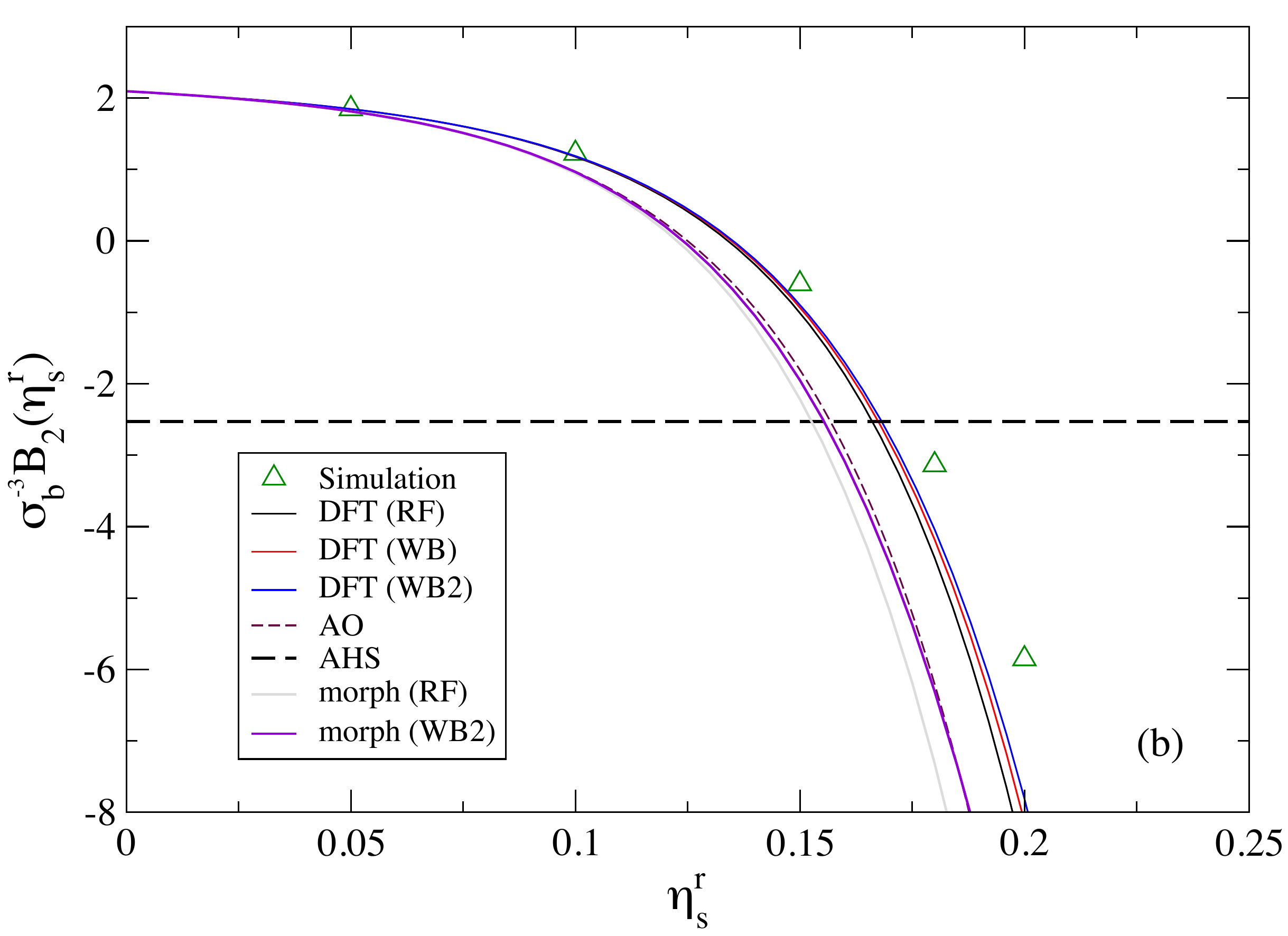}
\includegraphics[type=pdf,ext=.pdf,read=.pdf,width=0.85\columnwidth,clip=true]{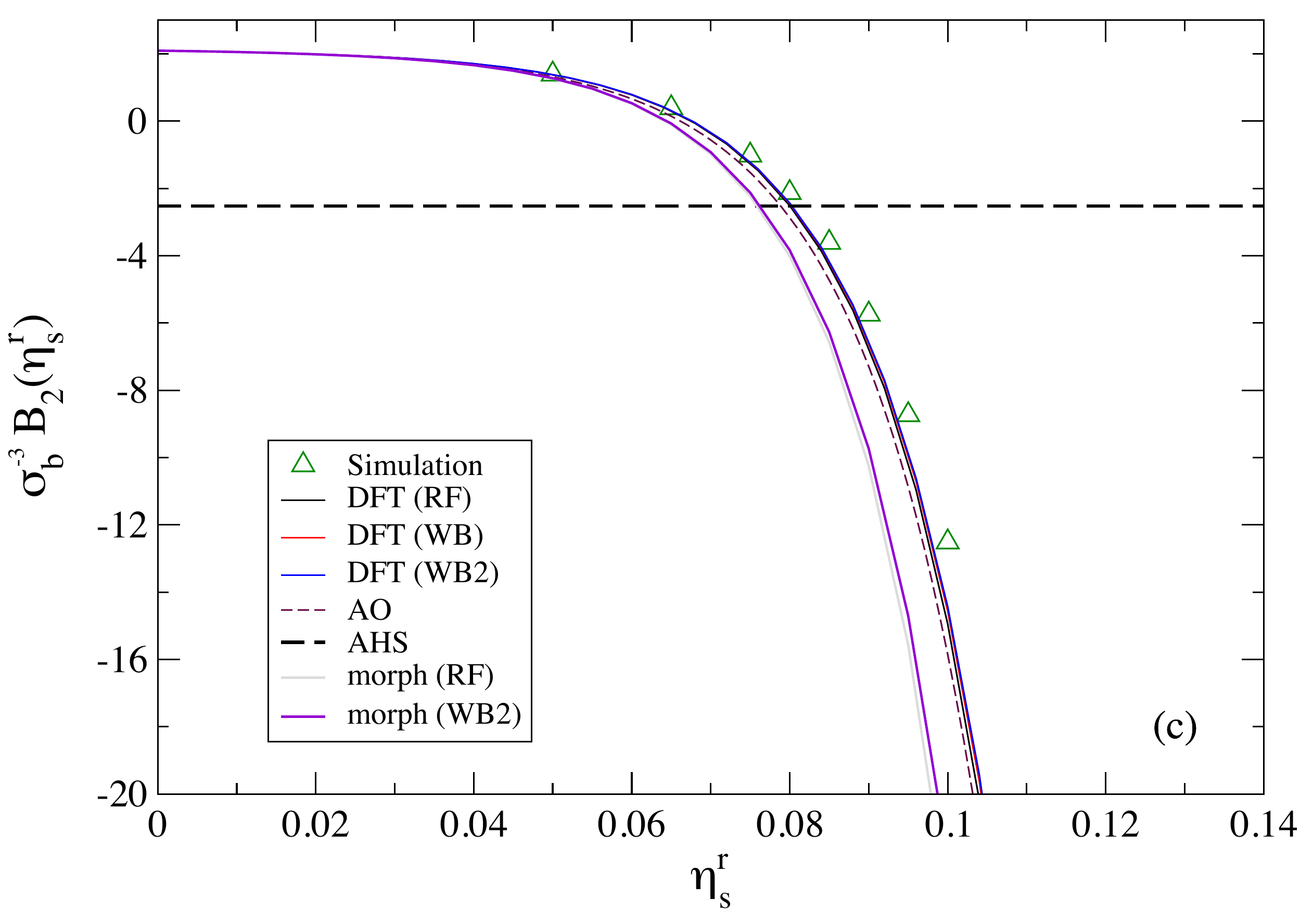}
\caption{{\bf (a)} Comparison of the second virial coefficient $B_2$
(Eq.~\protect\ref{eq:B2}) derived from DFT, morphometric and simulation measurements
of the effective potential for  $q=0.1$ and various $\eta_s^r$. The
horizontal dashed line indicates the value of $B_2^{AHS}=-2.527\sigma_b^3$ for
which fluid-fluid criticality was found in the adhesive hard sphere model
\cite{Miller2004,Largo2006}--see text. For values of $B_2$ below this
line, fluid-fluid phase separation is expected. {\bf (b)} and {\bf (c)}
show corresponding plots for $q=0.05$ and $q=0.02$ respectively.
Unless error bars are shown, statistical errors in simulation data points do not exceed the symbol size.}

\label{fig:B2}
\end{figure}

In practical terms the universality of $B_2$ at the fluid-fluid critical
point implies that one can predict the critical point value of
$\eta_s^r$ for effective one-component treatments of hard sphere
mixtures at small $q$ simply by matching the corresponding $B_2$ to the
universal value. Conversely, it follows that comparison of $B_2$ values
as a function of $\eta_s^r$ for different potentials provides a
sensitive measure of the extent to which their phase behaviour differs.
We have made such a comparison for effective potentials obtained from
DFT, the morphometric approximation and simulation for $q=0.1,0.05$ and
$0.02$. The results are shown in Fig.~\ref{fig:B2}(a-c) and demonstrate
that at the two larger values of $q$ even the relatively small
differences that we observe between the DFT and simulation estimates of
effective potentials could lead to significant differences in the small
particle packing fractions at which fluid-fluid phase separation is
predicted to occur. Specifically, for $q=0.1$ based on this $B_2$
criteria, it seems that the DFT with the Rosenfeld functional 
underestimates the putative critical point value of $\eta_s^r$ by some
$13\%$, while the WB2 functional underestimates it by some
$9\%$~\footnote{We note that in
Refs.~\protect\cite{Vliegenthart2000,Roth2001} an empirical (average)
value $B_2^{crit}=-1.5B_2^{HS}$ was used to estimate the critical point.
We prefer the AHS value as an indicator of the onset of phase separation
since we focus on short-ranged (sticky) potentials, following Largo and
Wilding \protect\cite{Largo2006}.} For $q=0.05$ (Fig.~\ref{fig:B2}(b))
the discrepancy between DFT and simulation has fallen to about $4\%$,
while for $q=0.02$ (Fig.~\ref{fig:B2}(c)), the values of $B_2$ for the
hard spheres mixtures arising from the various DFT flavors agree very
well with one another and with simulation, at least for the range of
$\eta_s^r$ at which bulk phase separation is expected to occur. They
also agree well with the AO model, suggesting that the additive and
extreme non additive models will have very similar phase behaviour at
this value of $q$. Recall that for $q<0.154$ the mapping of the binary
AO mixture to an effective one-component Hamiltonian, with the AO
depletion potential Eq.~\ref{eq:AOpot}, is exact and we might also
expect that for very small $q$  the phase behaviour of the full binary
hard sphere mixture, at physically relevant (small)  values of
$\eta_s^r$, is described accurately by the depletion pair potential we
calculate here. Many body contributions should be negligible. 

Interestingly, the DFT data for $B_2$ exhibit a broad minimum with
increasing $\eta_s^r$, as can be seen for $q=0.1$ in
Fig.~\ref{fig:B2}(a). The same feature has previously been reported in
Ref.~\cite{Roth2001}. A similar minimum occurs within the DFT for $q=0.05$ at
$\eta_s^r\approx 0.38$ (not shown in Fig.~\ref{fig:B2}(b)). The origin
of the upturn in the value of $B_2$ beyond the minimum appears to be
due to the fact that the magnitude of the first repulsive maximum of $W(r)$
increases faster with $\eta_s^r$ than the depth of the potential well at
contact. Unfortunately we could not corroborate the authenticity of this
feature via simulation because it occurs at larger values of $\eta_s^r$
than are currently accessible to us. Should it prove real (rather than
being an artifact of the DFT), it raises the intriguing possibility that
fluid-fluid phase separation may occur only within a certain range of
$\eta_s^r$.

Also plotted in Fig.~\ref{fig:B2} are the results of the morphometric
approximation Eq.~\ref{eq:fullmorph} for $B_2$. Like the DFT results
these show minima at all $q$ studied (though only that for $q=0.1$ is
visible in the plotted ranges). For $q=0.1$, $B_2(\eta_s^r)$ does not
cross the AHS line, implying that the theory fails to predict
fluid-fluid phase separation at this size ratio. At smaller $q$, the
agreement with simulation is better, but still poorer than DFT. It is
disappointing that both versions of the morphometric approximation
perform poorly for $q=0.02$, where we find the results are substantially
different from those of the AO model. The discrepancy with simulation
for $B_2$ appears to arise primarily from a failure of the morphometric
approximation to correctly predict the additive constant in the
potential, as shown from the comparison of the potentials of
Figs.~\ref{fig:q0.05-eta0.2} and \ref{fig:q0.1-eta0.35}  with the
shifted representation of Fig.~\ref{fig:derjag}. Whilst morphometric results for the depletion force
\cite{Oettel2009,Botan2009} might be in reasonable agreement with DFT
and simulation, any additive shift is important for $B_2$.

\section{Discussion}
\label{sec:discuss}

In summary, we have employed bespoke MC simulation techniques to obtain
direct and accurate simulation measurements of depletion potentials
in highly size asymmetrical binary mixtures of hard spheres having
$q\leq 0.1$. Small particles were treated grand canonically, the value
of the chemical potential being chosen to target prescribed values of
the reservoir packing fraction $\eta_s^r$. The simulation results were
compared with new DFT calculations (performed using the insertion
method) based on the Rosenfeld, White Bear and White Bear Mark 2
functionals. For $\eta_s^r\leq 0.2$ generally good agreement with
simulation was found at all size ratios studied, though on increasing
the packing fraction to $\eta_s^r=0.35$ at $q=0.1$ significant
discrepancies between the various flavors of DFT and the simulation
estimates were evident. In this latter regime, Rosenfeld (RF) was found to be
somewhat better than the other functionals in reproducing the height of
the first maximum of the effective potential, while White Bear 2 was
marginally the best of the three with regard to its prediction for the
contact value and for second virial coefficients. 

Overall our results show that the DFT insertion method provides a
reasonably accurate description of effective potentials for highly size
asymmetrical hard sphere mixtures at least in the range of small
particle packing fractions at which fluid-fluid phase separation is
likely to occur. Indeed at $\eta_s^r=0.35$, and $q=0.1$ DFT was found to
be more accurate than the morphometric and Derjaguin approximations, the
latter providing a particularly poor prediction. This conclusion is
partly at odds with that of Herring and Henderson
\cite{Herring06,Herring07} who assert that both DFT and the Derjaguin
approximation provide descriptions that are almost equally poor compared
to simulation data (see Fig.~7 of Ref.~\cite{Herring07} which refers to
$q=0.05$ and $\eta_s^r =0.3$ and $0.4$) and advocate in particular that
neither approach should be used in the regime of `moderate' $\eta_s^r$
to answer important questions such as the existence of fluid-fluid
coexistence. While we concur with this assessment in the case of the
Derjaguin approximation, Herring and Henderson's conclusions regarding
the accuracy of DFT calculations were reached on the basis of simulation
estimates for the effective potential which were not obtained directly,
but by integrating measurements of the interparticle force as outlined
in Sec.~\ref{sec:sims}. Perhaps as a consequence, their estimates are
much noisier (see Fig 6 of \cite{Herring07}) than those presented in the
present work and consequently --we feel-- do not serve as a sufficiently
reliably indicator of the accuracy of DFT especially in the key regime
where fluid-fluid phase separation might occur.

Indeed, we have investigated the likely extent of the consequences for
predictions of phase behaviour arising from discrepancies between theory
and simulation estimates of depletion potentials via calculations of the
dependence of the second virial coefficient on $\eta_s^r$. Previous
simulation studies of phase behaviour in single component fluid
interacting via effective potentials \cite{Largo2006} have shown that
when the potential is very short ranged, the onset of fluid-fluid phase
separation occurs at a near-universal value of $B_2=-2.527\sigma_b^3$.
Based on this criterion, we found that compared to the effective
potentials obtained via simulation in the present work, the morphometric
theory provides the poorest predictions of the critical packing fraction
of small particles (and fails to predict phase separation at all at
$q=0.1$). Those from DFT underestimate the critical packing fraction of
small particles by about $10\%$ for $q=0.1$ and about $4\%$ for
$q=0.05$. While these are significant discrepancies, we do not feel that
they constitute a ``qualitative breakdown'' of the DFT insertion method
approach as suggested by Herring and Henderson
\cite{Herring06,Herring07} on the basis of their simulation data, at
least not in the regime where phase separation is expected. Herring and
Henderson speak of a nanocolloidal regime. We interpret this as size
ratios $q$ of say $0.1$ to $0.01$. Our present study shows that this
regime is amenable to accurate simulation studies up to values of the
small sphere packing fraction that are relevant for investigations of
fluid-fluid phase separation and that DFT works well in this regime -the
focus of the present paper. For larger values of $\eta_s^r$ there are
serious issues concerning the accuracy of the existing DFT approaches
and we shed no new light on this interesting but somewhat extreme
regime.

Turning finally to the outlook for further work on highly size
asymmetrical mixtures, it would, of course, be of great interest to
verify the existence (or otherwise) of the putative fluid-fluid critical
point in the full two component size-asymmetric hard sphere mixture.
This topic remains ebullient. For example, a recent paper
\cite{Sillren:2010ly}, based on a version of thermodynamic perturbation
theory, conjectures that additive hard spheres will exhibit fluid-fluid
separation, albeit metastable with respect to the fluid-solid
transition, for size ratios in the range $0.01\le q\le 0.1$.  Our
measurements of $B_2$ for the depletion potentials obtained in our
simulations provide (potentially) accurate predictions for the small
particle packing fraction at which the critical point occurs. We are
currently employing a grand canonical version of the staged insertion MC
method \cite{Ashton:2011fk} to investigate this matter.

Our simulation methods can be applied to any short ranged potential and
it would also be of interest to examine the influence on the
effective potential of adding small amounts of finite attraction or
repulsion to the $bs$ and $ss$ interactions. This would allow us to
better model real colloidal systems, where one can have a variety of
interaction potentials, and where, even in systems (such as sterically
stabilized PMMA) which approximate hard spheres rather well, one expects
residual non-hard sphere interactions \cite{Germain2010}. Previous work
\cite{Louis2001,Roth2001} has suggested that the effects of such
residual interactions may be represented in terms of a {\em
non-additive} hard sphere mixture. It would be of interest to examine
this proposal explicitly using accurate simulation data and DFT
calculations. 

\acknowledgments

This work was supported by EPSRC grant EP/F047800 (to NBW). Simulation results
were partly produced on a machine funded by HEFCE's Strategic Research
Infrastructure fund. The authors are grateful to Martin Oettel for
comments on the original manuscript and in particular on the
morphometric approximation.

\bibliography{/Users/pysnbw/Dropbox/Papers}
\end{document}